\documentclass[a4paper,UKenglish,cleveref, autoref, thm-restate]{lipics-v2021}

\title{Stability Notions for Hospital Residents with Sizes\footnote{A preliminary version of this work appeared in the conference proceedings of FSTTCS, 2025~\cite{balasundaram_et_al_fsttcs_2025}.}}

\author{{Haricharan} {Balasundaram}}{Indian Institute of Technology Madras, India}{haricharanb@smail.iitm.ac.in}{https://orcid.org/0009-0009-8962-7734}{}
\author{{J} {B} {Krishnashree \textcolor{blue}{\footnote{Corresponding author.}}} \footnote{Part of this work was done during a semester-long project at IIT Madras, when the author was pursuing Integrated M.Sc., at PSG College of Technology, Coimbatore.}}{Tata Institute of Fundamental Research Mumbai, India}{krishnashree.j@tifr.res.in}{https://orcid.org/0009-0003-2287-7807}{}
\author{Girija Limaye \footnote{Part of this work was done when the author was a faculty at FLAME University, Pune and BITS Pilani, K K Birla Goa Campus, Goa.}}{Independent Researcher}{girija.iitm@gmail.com}{https://orcid.org/0009-0000-6241-6720}{}
\author{Meghana Nasre \textcolor{blue}{\footnote{Corresponding author.}}}{Indian Institute of Technology Madras, India}{meghana@cse.iitm.ac.in}{https://orcid.org/0000-0003-0290-4444}{}

\authorrunning{Balasundaram et al.} 

\Copyright{Haricharan Balasundaram and J B Krishnashree and Girija Limaye and Meghana Nasre} 



\category{} 

\acknowledgements{We thank the anonymous reviewers \textcolor{black}{of FSTTCS 2025} for their comments that significantly improved the presentation of the  paper.}

\nolinenumbers 

\usepackage{graphicx}
\usepackage{xcolor}
\usepackage{algorithm}
\usepackage[noend]{algpseudocode}
\usepackage{multicol}
\usepackage{chngcntr}
\usepackage{booktabs}
\usepackage{tabularx}
\usepackage{array}
\newcolumntype{Y}{>{\raggedright\arraybackslash}X}

\usepackage[title]{appendix}
\usepackage{tikz}
\usepackage{float}
\usetikzlibrary{decorations.pathreplacing}

\def\NP{NP}

\newcommand{\HRS}[0]{HRS}
\newcommand{\HR}[0]{HR}

\newcommand{\SMTI}[0]{SMTI}
\newcommand{\CSMTI}[0]{$(3,3)$-COM-SMTI}

\newcommand{\B}{\mathcal{H}}

\newcommand{\A}{\mathcal{A}}

\newcommand{\hide}[1]{}

\newcommand{\agent}[0]{agent}
\newcommand{\hospital}[0]{hospital}

\newcommand{\RNum}[1]{(\romannumeral #1)}

\def\smax{s_{\max}}

\def\rmax{r_{\max}}

\begin{document}
\fontsize{11}{13}\selectfont
\maketitle

\begin{abstract}
The Hospital Residents problem with sizes (\HRS{}) is a generalisation of the well-studied hospital residents (\HR) problem. In the \HRS{} problem, an agent $a$ has a size $s(a)$ and the agent occupies $s(a)$ many positions of the hospital $h$ when assigned to $h$. The notion of stability in this setting is suitably modified, and it is known that deciding whether an \HRS{} instance admits a stable matching is \NP-hard under severe restrictions. In this work, we explore a variation of stability, which we term occupancy-based stability. This notion was defined by McDermid and Manlove (J. of Comb. Opt. 2010) but remained unexplored to the best of our knowledge. In our work, we show that every \HRS{} instance admits an occupancy-stable matching. We further show that computing a maximum-size occupancy-stable matching is \NP-hard, even in the presence of master list ordering. We complement our hardness result by providing an approximation algorithm with a guarantee strictly better than $3$ for the max-size occupancy-stable matching problem.

Given that the classical notion of stability adapted for \HRS{} is not guaranteed to exist in general, we show a practical restriction under which a stable matching is guaranteed to exist. We present an efficient algorithm to output a stable matching in the restricted \HRS{} instances. We also provide an alternate \NP-hardness proof for the decision version of the stable matching problem for \HRS{} which imposes a severe restriction on the number of neighbours of non-unit sized agents.
\end{abstract}

\section{Introduction}\label{sec:intro}

The Hospital Residents (\HR) setting is a well-studied scenario in the literature where a set of agents, commonly called as a set of residents, needs to be assigned to hospitals. Each agent strictly orders the set of hospitals acceptable to the agent. This ordering is called the preference list. Similarly, each hospital also has a strict ordering of agents acceptable to the hospital. In addition, each hospital has a capacity that denotes the maximum number of agents that can be assigned to the hospital. The goal in the \HR\ setting is to compute an assignment of agents to hospitals which respects the capacities and is {\em optimal} with respect to the preferences submitted. The \HR\ setting models several important applications, namely, recruiting medical interns (residents) in hospitals~\cite{nrmp,gale_shapley}, admitting students to degree programs; for example, the Joint Seat Allocation Authority (JOSAA) that is used in undergraduate admissions in India~\cite{josaa,baswana2019centralized}, and many similar scenarios~\cite{roth,HR_example_1}. In most of these settings, the de facto standard of optimality with preferences is the notion of stability. Informally, an assignment is stable if no agent hospital pair wishes to deviate from the assignment. It is well known that every instance of the \HR\ setting admits a stable assignment~\cite{gale_shapley}.

Since the \HR\ setting is relevant in various practical scenarios, it has been extensively studied in the literature. Furthermore, several generalisations of the \HR\ setting are investigated. In this paper, we consider a particular variant in which agents can have sizes, and we denote this setting as the \HRS\ setting (\HR\ with (agents having) sizes) throughout the paper. An agent $a$ in the \HRS\ setting has a positive size $s(a)$ associated with it. This agent can be thought of as a group of residents/students, all of whom have the same preference ordering and have to be assigned together in any assignment. Furthermore, assigning an agent $a$ of size $s(a)$ to a hospital $h$ occupies $s(a)$ many positions at $h$.

As mentioned in~\cite{Refugee, MatchingMarket_with_Sizes}, the \HRS\ setting models multiple applications which include refugee settlement and accommodating children in day-cares. In the refugee settlement problem refugee families require multiple units of services (say, hospital beds) from each host locality. The refugee family can be represented as an agent with size proportional to the demand of the family, and the host locality with its service capacity can be represented by a hospital. A similar scenario arises in accommodating children in day-care centres. It can be realised where it is desirable to have siblings in the same day care or every child occupies a certain number of slots in the day care per day. The siblings (or the number of slots needed per child) can be represented by an agent of a given size, and day cares typically have a capacity representing how many children can be accommodated (or available slots). These correspond to hospitals in our setting.

In the presence of sizes for agents, the notion of stability is suitably adapted to take into account agent sizes (see Definition~\ref{def:bp} and Definition~\ref{def:stable_hrs}). It is well known that an \HRS\ instance may fail to admit a stable matching (see Fig.~\ref{fig:ex1}).  Furthermore, deciding whether an instance of the \HRS\ setting admits a stable matching is \NP-hard~\cite{hrs_hardness}. Thus, the natural extension of the notion of stability does not offer a practical solution. In the work of McDermid and Manlove~\cite{hrs_hardness}, which investigated the \HRS{} problem, the authors had defined an alternate notion of stability, which we call occupancy-based stability (see Definition~\ref{deF:stronger_bp} and Definition~\ref{def:weak_stable}). We show that occupancy-stable matching always exists and can be computed efficiently.

As discussed in~\cite{hrs_hardness}, the rationale behind such an extended definition is as follows: in order to accommodate a better-preferred agent, the hospital may have to displace a set of lower-preferred agents. However, this displacement may create more capacity at the hospital than what is needed for the better-preferred agent. The hospital may prefer being more occupied over getting better-preferred agents. The alternate notion of stability ensures that this requirement is fulfilled. This notion was defined in the concluding remarks in \cite{hrs_hardness} but has not received attention to the best of our knowledge.

In our work, we investigate the stability notion adapted for \HRS\ (henceforth termed as stability) as well as the notion of occupancy-stable matchings. We show that every instance of the \HRS\ admits an occupancy-stable matching, and such a matching can be computed efficiently. We further show that the computation of maximum size occupancy-stable matching is \NP-hard. To complement this hardness, we provide an efficient algorithm with an approximation ratio strictly better than $3$ for the maximum size occupancy-stable matching in an \HRS\ instance. Furthermore, to circumvent the hardness of the decision version of the stable matching problem for \HRS\ instances, we investigate restricted cases. Master lists in preferences is a well-studied restriction \cite{master_list} where preferences of agents in a set are derived from a common ranking of the agents of the other side of the bipartition. We show that under a natural generalised master list restriction, every instance of the \HRS\ setting admits a stable matching which can be efficiently computed. We establish an analogue of the well-known Rural Hospitals Theorem for the \HRS\ setting.

\subsection{Notation and problem definition}
Let $G = (\mathcal{A} \cup \mathcal{H}, E)$ be a bipartite graph, with $n = |\mathcal{A} \cup \mathcal{H}|$ and $m = |E(G)|$. Let $\mathcal{A}$ denote the set of agents and $\mathcal{H}$ denote the set of hospitals. An edge $(a, h) \in E$ indicates that $a$ and $h$ are mutually acceptable. For each vertex $v \in \mathcal{A} \cup \mathcal{H}$, we let $\mathcal{N}(v)$ denote the set of acceptable partners of $v$, that is, the set of vertices from the other set that are adjacent to $v$ in $G$. Each agent $a$ $\in \mathcal{A}$ has an integral positive size associated with it, denoted as $s(a)$. Let $ \smax = \max_{a \in \mathcal{A}} s(a)$. Every hospital $h$ has an integral positive capacity $q(h)$ associated with it, which denotes the maximum total size of agents that it can be assigned to.
An agent $a$ is called a non-unit sized agent if $s(a) > 1$.

Each agent $a$ ranks hospitals $\mathcal{N}(a)$ in a strict order, denoted as the preference list of $a$. Similarly, each hospital $h$ ranks agents in $\mathcal{N}(h)$ in a strict order, denoted as the preference list of $h$. For any vertex $v \in \mathcal{A} \cup \mathcal{H}$, we say that $v$ prefers $x$ over $y$ iff $x$ is before $y$ in $v$'s preference order, and we denote this by $x \succ_v y$ (see the example in Fig.~\ref{fig:ex1}, discussed later). Let $\ell_a$ (resp. $\ell_h$) denote the length of the longest preference list of an agent (resp. hospital). We abuse the notation and call the graph $G=(\mathcal{A} \cup \mathcal{H}, E)$ along with preferences, sizes and capacities an \HRS\ instance. It is easy to see that the \HR\ instance~\cite{gale_shapley} is a special case of \HRS\ instance with $\smax = 1$.

A many-to-one matching (called matching here onwards) in $G$ is an assignment between agents and their acceptable hospitals such that every agent is matched to at most one hospital and for every hospital $h$, the sum of the sizes of agents matched to $h$ is less than or equal to its capacity, that is, $q(h)$. Let $M(a)$ denote the hospital matched to agent $a$ in matching $M$, and $M(h)$ denote the set of agents matched to hospital $h$ in matching $M$. We assume that every agent prefers to be matched to one of its acceptable partners over being unmatched. The occupancy of $h$ w.r.t. matching $M$ is the total size of agents matched to $h$ in $M$, denoted by $O_M(h)$. Thus, $O_M(h) = \sum_{a\in M(h)}{s(a)}$.

\vspace{0.1in}
\noindent {\underline{\em Stable Matching:}} In an \HR\ instance, an agent-hospital pair $(a,h)$ is said to {\em block} a matching $M$, if $(a,h) \notin M$ and they both have an incentive to deviate from $M$ and instead, get matched to each other. That is, either $a$ is unmatched or $h \succ_a M(a)$, and either $O_M(h) < q(h)$ or there exists an agent $a' \in M(h)$ and $a \succ_h a'$.
A matching $M$ is stable if no pair blocks it.
The notion of blocking pair and stability can be extended for the \HRS\ setting as follows~\cite{hrs_hardness}.

\begin{definition}[Blocking pair]\label{def:bp}
Given a matching $M$ in an \HRS\ instance $G$, a pair $(a, h) \in E \setminus M$ is a blocking pair w.r.t. $M$ if $h \succ_a M(a)$ and there exists a set of agents $X \subseteq M(h)$, possibly empty, such that for every $a' \in X$, $a \succ_h a'$ and  $O_M(h) - \sum_{a' \in X} s(a') + s(a) \le q(h)$.
\end{definition}

The definition implies that an unmatched edge $(a, h)$ is a blocking pair if $a$ is either unmatched or prefers $h$ to its current matched partner in $M$, and $h$ has sufficient capacity to accommodate $a$ by replacing a set $X$ of agents that are currently matched to $h$ and are lower-preferred over $a$. Note that if $X$ is empty, it implies that $h$ can accommodate $a$ without replacing any agent(s) in $M(h)$.

\begin{definition}[Stable matching]\label{def:stable_hrs}
    A matching $M$ in an \HRS\ instance $G$ is stable if there does not exist a blocking pair w.r.t. $M$.
\end{definition}

It is well-known that an \HR\ instance always admits a stable matching. The celebrated algorithm by Gale and Shapley efficiently computes a stable matching in an \HR\ instance~\cite{gale_shapley}. In contrast to this, an \HRS\ instance may not admit a stable matching, as illustrated in the following example from~\cite{hrs_hardness}. Fig.~\ref{fig:ex1} shows an example instance with $\mathcal{A} = \{a_1, a_2, a_3\}$ and $\mathcal{H} = \{h_1, h_2\}$. The preference lists of agents and hospitals are as shown in the Fig.~\ref{fig:ex1}. We observe that any matching that leaves $a_1$ unmatched cannot be stable since $(a_1, h_1)$ blocks the matching. Similarly, any matching that leaves $a_2$ unmatched cannot be stable since $(a_2, h_2)$ blocks the matching. Therefore, if a stable matching exists in the instance, it must match {\em both} $a_1$ and $a_2$. Irrespective of where $a_1$ and $a_2$ are matched, the capacity constraint does not allow $a_3$ to be matched. Thus, we consider $M = \{(a_1,h_2),(a_2,h_2)\}$, $M'=\{(a_1,h_2),(a_2,h_1)\}$ and $M''=\{(a_1,h_1),(a_2,h_2)\}$. It is easy to verify that $M$ is blocked by $(a_2, h_1)$,  $M'$ is blocked by $(a_3, h_2)$ and $M''$ is blocked by $(a_1, h_2)$. Thus, the instance does not admit a stable matching.

\begin{figure}
\begin{align*}
   (1) \ a_1 &: h_2 \succ h_1 & (1)\  a_2 &: h_1 \succ h_2 &  (2) \ a_3 &: h_2 \\
   [1] \ h_1 &: a_1 \succ a_2 & [2] \ h_2 &: a_2 \succ a_3 \succ a_1
\end{align*}
\caption{An \HRS\ instance that does not admit a stable matching. The number preceding the agent shows the size of the agent, whereas the number preceding the hospital denotes the hospital capacity. For example, $s(a_3) = 2$ and $q(h_2) = 2$. The preference list of an agent or a hospital is as shown. For example, hospital $h_2$ prefers agent $a_2$, followed by agent $a_3$, followed by agent $a_1$.}
\label{fig:ex1}
\end{figure}

McDermid and Manlove~\cite{hrs_hardness} showed that the problem of determining whether an \HRS\ instance admits a stable matching is \NP-hard. In light of this, we consider restricted settings of the instance under which the problem becomes tractable. One such restriction that has been investigated is the assumption of {\em master list}. The set of agents is said to have a master list ordering on them if there exists an ordering of agents such that every hospital has its preference list derived from that ordering~\cite{master_list}. We propose and investigate the notion of a  {\em generalised master list} as defined below. Unlike an arbitrary  \HRS\ instance which may fail to admit a stable matching, we show that an \HRS\ instance with the generalised master list ordering always admits a stable matching, and it can be computed efficiently. 

\smallskip

\noindent{\em Generalised master list on agents.} We say that the preference lists of hospitals follow the generalised master list ordering on agents if there exists a partition $\{\mathcal{A}_{t_1}, \mathcal{A}_{t_2}, \ldots, \mathcal{A}_{t_f}\}$ of $\mathcal{A}$ such that all agents in $\mathcal{A}_{t_i}$ have the same size and there exists an ordering $\langle \mathcal{A}_{t_1}, \mathcal{A}_{t_2}, \ldots, \mathcal{A}_{t_f} \rangle$ on the sets in the partition of $\mathcal{A}$ such that every hospital $h \in \mathcal{H}$ prefers its neighbours in $\mathcal{A}_{t_i}$ over its neighbours in $\mathcal{A}_{t_j}$ iff $i < j$.  We note that generalised master list ordering does {\em not} impose any master list ordering within a fixed set $\mathcal{A}_{t_i}$ of the partition. Next, we illustrate generalised master lists via an example.

Let $\mathcal{A} = \{a_1, a_2, a_3, a_4, a_5\}$ and  $\mathcal{H} = \{h_1, h_2, h_3\}$. The agent preferences and the hospital capacities do not matter for the illustration and hence are left unspecified.  The preferences of the hospitals, the relationship between the sizes of the agents and a partition of the agent set are shown in Fig.~\ref{fig:ex2}. It is noted here that in the example, $s(a_1)$ and $s(a_4)$ could be equal.  We also remark that although the generalised master list is defined on all agents in $\mathcal{A}$, the individual hospitals may have an incomplete preference list as seen in the example.

\begin{figure}[ht]
\begin{align*}
    h_1 &: a_2 \succ a_1 \succ a_4 & h_2 &: a_1 \succ a_2 \succ a_3 & h_3 &: a_5 \succ a_4 
\end{align*}

\caption{Illustration of generalised master lists. Assume that 
$s(a_1) = s(a_2) \neq s(a_3)$ and $s(a_4) = s(a_5) \neq s(a_3)$. Let $\mathcal{A} = \{\mathcal{A}_{t_1}, \mathcal{A}_{t_2}, \mathcal{A}_{t_3}\}$ where  $\mathcal{A}_{t_1} = \{a_1, a_2\}, \mathcal{A}_{t_2} = \{a_3\}, \mathcal{A}_{t_3} = \{a_4, a_5\}$.}
\label{fig:ex2}
\end{figure}

It is easy to see that the preferences of hospitals follow the generalised master list ordering on agents because there exists an ordering $\langle \mathcal{A}_{t_1}, \mathcal{A}_{t_2}, \mathcal{A}_{t_3} \rangle$ on the sets in the partition such that every hospital $h$ prefers all agents in $\mathcal{A}_{t_i} \cap \mathcal{N}(h)$ over $\mathcal{A}_{t_j}  \cap \mathcal{N}(h)$ {iff} $i < j$. Since the generalised master list ordering does {\em not} impose any ordering within a fixed set $\mathcal{A}_{t_i}$, we note that in Fig.~\ref{fig:ex2} both $a_1, a_2$ belong to $\mathcal{A}_{t_1}$ and yet it is acceptable that $a_2 \succ_{h_1} a_1$ and $a_1 \succ_{h_2} a_2$. 
We remark that the example in Fig.~\ref{fig:ex1} does not follow a generalised master list since there is no way to partition the three agents satisfying the required properties.
We use generalised master lists in the context of the standard stability notion.

We now turn to an alternative notion of stability, which is a relaxation of the standard notion and is appealing since we show that it ensures guaranteed existence.

\noindent \underline{\em {Occupancy Stable Matching:}} McDermid and Manlove~\cite{hrs_hardness} discuss the following notion of a blocking pair, which is stronger than the one in Definition~\ref{def:bp}. We term it an occupancy-blocking pair.

\begin{definition}[Occupancy-blocking pair]\label{deF:stronger_bp}
    Given a matching $M$ in an \HRS\ instance $G$, a pair $(a, h) \in E \setminus M$ is a blocking pair w.r.t. $M$ if $h \succ_a M(a)$ and there exists a set of agents $X \subseteq M(h)$, possibly empty, such that $\forall a' \in X$, $a \succ_h a'$ and  $\displaystyle O_M(h) - \sum_{a' \in X} s(a') + s(a) \le q(h)$ and $s(a) \geq \displaystyle  \sum_{a' \in X} s(a')$.
\end{definition}

That is, an edge $(a,h)$ is an occupancy-blocking pair w.r.t. matching $M$ if it satisfies Definition~\ref{def:bp} and the occupancy of $h$ does not reduce by replacing the agents in $X$ with agent $a$. When a matching $M$ does not admit any occupancy-blocking pair, then $M$ is an occupancy-stable matching.

\begin{definition}[Occupancy-stable matching]\label{def:weak_stable}
    A matching $M$ in an \HRS\ instance $G$ is occupancy-stable if there does not exist an occupancy-blocking pair w.r.t. $M$.
\end{definition}

 For the \HRS\ instance in Fig.~\ref{fig:ex1} that does not admit any stable matching, consider the matching $N = \{(a_1, h_1), (a_3, h_2)\}$. The pair $(a_2,h_2)$ is not an occupancy-blocking pair since $s(a_3) > s(a_2)$, hence $N$ is occupancy-stable. As mentioned earlier, in this work we show that every instance of the \HRS\ setting admits an occupancy-stable matching and it can be efficiently computed. 

{The size of a matching in an \HRS\ instance is the sum of the sizes of the matched agents, or equivalently, the size of a matching is the total occupancy of all hospitals. There are simple instances in which different occupancy-stable matchings have different sizes (see the example in Fig.~\ref{fig:ex_2bound}). Hence, a natural question is to ask for an occupancy-stable matching which maximises the size. In this work, we show that computing such a matching is \NP-hard. We complement it with an approximation algorithm with guarantee strictly better than $3$.

\noindent \underline{\em First Coalition Stable Matching:} We now study another variant of the standard stability notion, which strengthens occupancy-stability and relaxes the notion of stability. The notion was introduced by Rodríguez and Manlove~\cite{manlove_journal} in the context of Course Allocation with Credits, which is adapted to \HRS . Section~\ref{sec:related_work} discusses in detail how \HRS\ can be viewed as a restriction of the course allocation with credits setting, and how the results of \cite{manlove_journal} carry over to the \HRS\ setting. Now, observe that an occupancy-blocking pair allows only a single agent to replace a set of lower-ranked agents such that its size is large enough. However, a hospital may benefit from simultaneously accepting multiple agents whose combined size is sufficiently large. This motivates the notion of a first-blocking coalition.

\begin{definition}[First-blocking coalition]\label{def:fc_blocking}
Let $M$ be a matching in an \HRS{} instance $G$. A hospital $h$ and a set of agents $B \subseteq \mathcal{N}(h) \setminus M(h)$ form a first-blocking coalition w.r.t.~$M$ if the following conditions hold.
\begin{itemize}
\item For every $a \in B$, either $a$ is unmatched or $h \succ_a M(a)$, and
\item There exists a set $D \subseteq M(h)$ such that the following holds simultaneously:
\begin{itemize}
    \item Let $a$ denote the most preferred agent in $B$ by $h$ and let $d$ denote the most preferred agent in $D$ by $h$. Then  $a \succ_h d$.
    \item \( \sum_{a \in B} s(a) \geq \sum_{a' \in D} s(a').\)
    \item\(
    O_M(h)-\sum_{a' \in D}s(a')
    +\sum_{a \in B}s(a)
    \le q(h).
    \)
\end{itemize}
\end{itemize}
\end{definition}

The definition implies that all agents in the coalition strictly prefer the hospital $h$ to their assignment in $M$. Furthermore, the hospital can accommodate the coalition by replacing a set of currently assigned agents. The occupancy of the hospital does not decrease after the replacement, and the most preferred new agent is preferred over the most preferred replaced agent. Moreover, suppose there exists a first-blocking coalition with a set of agents $B$ and a hospital $h$ such that $D=\emptyset$. For any agent $a \in B$, the pair $(a,h)$ is an occupancy blocking pair, as $h$ can accommodate $a$ without unmatching any currently matched agents. Furthermore, when both $B$ and $D$ are singleton sets, a first-blocking coalition is precisely an occupancy blocking pair. Proposition~\ref{prop:sfcpsm} states the relation between the three stability notions.

\begin{definition}[First-coalition stable matching]
A matching $M$ in an \HRS{} instance $G$ is first-coalition stable if there does not exist a first-blocking coalition w.r.t.~$M$.
\end{definition}

We note that the instance in Fig.~\ref{fig:ex1} does not admit a first-coalition stable matching, as proved in~\cite{manlove_journal}.

In Fig.~\ref{fig:sm_fcsm} we give an \HRS\ instance which does not admit any stable matching: this is because, any stable matching (if it exists) must match $(a_4,h_3)$. This leaves us with an instance equivalent to an instance that does not admit any stable matching, as seen in Fig.~\ref{fig:ex1}. But this instance admits a first-coalition stable matching $M=\{(a_1,h_1),(a_2,h_2),(a_3,h_3)\}$. 

\begin{figure}[!ht]
\centering
\begin{align*}
(1)\ a_1 &: h_1 \succ h_2 &
(1)\ a_2 &: h_2 \succ h_1 &
(2)\ a_3 &: h_2 \succ h_3 &
(1)\ a_4 &: h_3  \\
[1]\ h_1 &: a_1 \succ a_2 &
[2]\ h_2 &: a_2 \succ a_3 \succ a_1 & 
[2]\ h_3 &: a_4 \succ a_3
\end{align*}

\caption{An \HRS\ instance that does not admit any stable matching but admits a first-coalition stable matching. 
}
\label{fig:sm_fcsm}
\end{figure}

\begin{proposition}
Every stable matching is first-coalition stable and every first-coalition stable matching is occupancy-stable.
\label{prop:sfcpsm}
\end{proposition}

\begin{proof}
    Note that if there is an occupancy-blocking pair $(a,h)$, then it is also a first-blocking coalition (Definition~\ref{def:fc_blocking}) where $B=\{a\}$ with $h$ and also the classical blocking pair (Definition~\ref{def:bp}). And if there is a first-blocking coalition $B$ and $h$, then $(a,h)$ where $a$ is the most preferred agent in $B$ is a classical blocking pair. Immediately, no classical blocking pair implies no first-blocking coalition and no occupancy-blocking pair. 
\end{proof}

\subsection{Our results} \label{sec:our_results}
We show the following new results for the \HRS\ setting. 

\smallskip

\noindent{\bf Occupancy-stable matchings.}
We show that every \HRS\ instance admits an occupancy-stable matching, and such a matching can be efficiently computed. However, the problem of computing a maximum size occupancy-stable matching is \NP-hard.

\begin{theorem}\label{thm:occ_stable_existence}
An \HRS\ instance always admits an occupancy-stable matching, and it can be computed in $O(m+n\log{n})$ time.
\end{theorem}

\begin{theorem}\label{thm:occ_stable_hardness}
    Given an \HRS\ instance, computing a maximum size occupancy-stable matching is \NP-hard, even if the size of each agent is at most 3, and there is a master list ordering on both agents and hospitals.
\end{theorem}

In the conference version of our work~\cite{balasundaram_et_al_fsttcs_2025}, we showed the \NP-hardness even if the size of each agent and the capacity of each hospital is at most 2, and the lengths of the agents’ and hospitals’ preference lists are at most 4.
Note that neither this result nor Theorem~\ref{thm:occ_stable_hardness} strengthen the other. We also note that the \NP-hardness of computing a max-size occupancy stable matching is also implied by Theorem~{5.11} in~\cite{manlove_journal} but our reduction is simpler as compared to the reduction presented in~\cite{manlove_journal}.

On the positive side, we show that there exists an approximation algorithm for computing a maximum size occupancy-stable matching with guarantee strictly better than $3$.
Let $s_1, s_2, \ldots, s_k$ denote the distinct sizes of agents in $G$ and let $s_1 > s_2 > \ldots > s_k \geq 1$.
Let $j \geq 2$. Thus, $s_j$ is any size except the largest size $s_1$.
Then $s_{j-1}$ is the immediate next size larger than $s_j$.
Let $\rmax$ be the ratio $\frac{(s_j) - 1}{s_{j-1}}$.

\begin{theorem}\label{thm:occ_stable_approx}
Given an \HRS\ instance, an occupancy-stable matching with an approximation guarantee of $\left(2+\rmax\right)$ can be computed in polynomial time.
\end{theorem}

Since $s_{j-1} > s_j > (s_j)-1$, we have that $\rmax < 1$.
Therefore the guarantee in Theorem~\ref{thm:occ_stable_approx} is strictly better than $3$.
Theorem~\ref{thm:occ_stable_approx} has the following corollary.
\begin{corollary}
    Given an \HRS\ instance with two sizes, $s_1 > s_2$ such that $s_2 = 1$, an occupancy-stable matching with guarantee $2$ can be computed in polynomial time.
\end{corollary}

In the conference version of this work~\cite{balasundaram_et_al_fsttcs_2025}, we presented an approximation guarantee of $3-\frac{1}{s_1}$. We significantly improve this guarantee in the current work to $2+\rmax$. Our improved analysis combines many crucial observations about the sizes of agents that are left unmatched in our algorithm's output but matched in an optimal matching. These observations combined together in a careful manner give us the improved guarantee.

\smallskip
\noindent{\bf Stable matchings.}
We show that when an \HRS\ instance admits a generalised master list ordering on agents, a stable matching always exists and it can be efficiently computed.

\begin{theorem}\label{thm:gen_master_list}
Given an \HRS\ instance with a generalised master list on agents, a stable matching always exists and it can be computed in $O(m+n)$ time.
\end{theorem}

In the classical \HR\ setting, the Rural Hospital's Theorem~\cite{Gus}  shows properties that remain  invariant across all stable matchings in an \HR\ instance. 
{We establish an analogue of the Rural Hospital's Theorem for HRS instance with generalized master list on agents. 
}

\begin{theorem}
\label{thm:max_size_gen_master_list}
Given an \HRS\ instance with a generalised master list on agents, all stable matchings match the same set of agents and hence all stable matchings are of the same size.
\end{theorem}

As a consequence of the above, in an HRS instance with generalized master list on agents our algorithm computes a maximum size stable matching. We remark that {\em without }a generalized master list assumption, Theorem~\ref{thm:max_size_gen_master_list} does not hold as illustrated by the example in Figure~\ref{fig:diff_size_sm}.
\begin{figure}[!ht]
\centering
\begin{align*}
   (3)\ a_1 &: h_1 \succ h_2 &
   (2)\ a_2 &: h_1 &
   (2)\ a_3 &: h_2 \succ h_1\\
   [4]\ h_1 &: a_3 \succ a_1 \succ a_2 &
   [3]\ h_2 &: a_1 \succ a_3
\end{align*}
\caption{An \HRS\ instance without generalized master list. The instance admits two stable matchings $M_1=\{(a_1,h_1),(a_3,h_2)\}$
with size $5$, and $M_2=\{(a_2,h_1),(a_3,h_1),(a_1,h_2)\}$ with size $7$.}
\label{fig:diff_size_sm}
\end{figure}

We note that the following are special cases of the generalised master list setting. Therefore, our algorithmic results (Theorem~\ref{thm:gen_master_list} and Theorem~\ref{thm:max_size_gen_master_list}) apply to them.

\begin{itemize}
    \item \textbf{Master list on agents}: Suppose there is a master list ordering on agents as $a_1' \succ a_2' \succ a_3'\succ \ldots$. Let $\mathcal{A}_{t_i} = \{a_i'\}$. Then this is a special case of the generalised master list ordering with the ordering $\langle \mathcal{A}_{t_1}, \mathcal{A}_{t_2}, \mathcal{A}_{t_3}, \ldots\rangle$.
\end{itemize}

Suppose that the distinct sizes of agents are $\{s_1, s_2, \ldots, s_k\}$, where $s_i > s_j$ whenever $i < j$. Let $\mathcal{A}_{s_i}$ indicate the set containing all agents with size $s_i$. 

\begin{itemize}

    \item \textbf{All agents with larger size are preferred over all agents with smaller size:} This is a special case of the generalised master list ordering with the order $\langle \mathcal{A}_{t_1} = \mathcal{A}_{1}, \dots, \mathcal{A}_{t_f} = \mathcal{A}_{s_k}\rangle$

    \item \textbf{All agents with a smaller size are preferred over all agents with a larger size:} This is a special case of the generalised master list ordering with the order $\langle \mathcal{A}_{t_1} = \mathcal{A}_{s_k}, \dots, \mathcal{A}_{t_f} = \mathcal{A}_{1}\rangle$
\end{itemize}

\smallskip
\noindent{\bf First-coalition Stable Matchings.} 
Our next theorem gives an alternate proof for the NP-hardness of the decision version of the first coalition stable matching problem in \HRS\ instances. The NP-hardness for the problem is already established by Theorem~4 in \cite{manlove_journal} but neither of the hardness results imply the other.

\begin{theorem}
    {For an \HRS\ instance, deciding whether it admits a first-coalition stable matching is \NP-hard even when each agent with non-unit size has exactly one hospital in its preference list.}
    \label{thm:nonunit_fc_Reduction}
\end{theorem}

Theorem~\ref{thm:nonunit_fc_Reduction} and proposition \ref{prop:sfcpsm} immediately imply the below corollary decision version of stable matching in \HRS\ setting. }

\begin{corollary}
    For an \HRS\ instance, deciding whether it admits a stable matching is \NP-hard even when each agent with non-unit size has exactly one hospital in its preference list.
\end{corollary}

The \NP-hardness for the decision version of stable matching in \HRS\ is proved in~\cite{hrs_hardness}. The hardness established in \cite{hrs_hardness} holds even when the degree of agents is bounded by three.  In our reduction, all non-unit sized agents have degree exactly one.
\subsection{Related work}
\label{sec:related_work}

The \HRS\ problem is related to the hospital residents problem with couples, where certain pairs of residents are identified as a couple~\cite{hrs_hardness}. The couples are allowed to give individual preferences as well as joint preferences. When the preferences of the couples are consistent, then the restriction is termed as Hospital residents with consistent couples. When a couple needs to be assigned to the same hospital, then they are called inseparable couples~\cite {hrs_hardness}. The \HRS\ problem considered here is a generalisation of hospital residents with consistent and inseparable couples. Our work builds on the work done by McDermid and Manlove~\cite{hrs_hardness} where they considered the \HRS\ problem. They showed that determining whether a stable matching exists in an \HRS~instance is \NP-hard, even when each preference list has length at most three and hospital capacities are at most two. 

The work of Delacretaz~\textit{et al.}~\cite{Refugee} considered a richer model where agents can have multidimensional constraints, which is a vector of demands from multiple services provided by the hospital. In the 2019 work by Delacretaz~\cite{MatchingMarket_with_Sizes}, the author worked with a restricted version of the HRS problem where only two different sized agents were considered. Since deciding whether a stable matching exists is \NP-hard, other fairness notions are considered. Delacretaz~\cite{MatchingMarket_with_Sizes} characterised a stable matching, if one exists, using these fairness notions. 

\HRS{} setting is also relevant to job assignment settings where each job has a size or processing time. Dean~\textit{et al.}~\cite{dean2006unsplittable} studied the problem of assigning jobs to machines with preferences, generally known as the ``ordinal'' assignment problem. The model they studied is the same as the \HRS\ setting, but their definition of feasible matching is relaxed. They allowed a small violation of hospital capacity. They proposed a polynomial-time algorithm that computes an unsplit (integral) job-optimal stable assignment in which each machine is overcapacitated by at most the processing time of the least-preferred job. In our work, we do not permit capacities to be violated, but instead consider a different notion of stability.

Rodr\'{i}guez and Manlove~\cite{manlove_journal} studied a generalisation of the \HRS\ problem in the context of course allocation with credits, where residents are analogous to courses and hospitals are analogous to students. In their model, each course has an associated credit value and quota, where the quota denotes the maximum number of students that can be assigned to the course, and each student has a credit limit. Unlike in \HRS, a course may be assigned to multiple students, analogous to allowing a resident to be matched to multiple hospitals. Thus, the \HRS\ setting is a special case of course allocation with credits in which every course has quota one. They studied four notions of stability, namely pair stability (corresponds to classical stability), pair-size stability (corresponds to occupancy stable matching as defined above), coalition stability, and first-coalition stability. Since their hardness reductions assume that every course has quota one, all of the results they establish for these notions apply directly to the \HRS\ setting. Furthermore, the complexity results for classical stability, first-coalition stability, and occupancy stability in the \HRS\ problem are summarised in Table~\ref{tab:complexity}, together with the contributions of this paper.

\begin{table}[t]
\centering
\small
\renewcommand{\arraystretch}{1.15}

\begin{tabularx}{\textwidth}{>{\raggedright\arraybackslash}p{3.2cm}YYY}
\toprule
\textbf{Problem}
& \textbf{Stable}
& \textbf{First-coalition}
& \textbf{Occupancy} \\
\midrule

Find
&
NP-h~\cite[Thm.~3.8]{hrs_hardness}, Thm.~\ref{thm:nonunit_fc_Reduction}
&
NP-h~\cite[Thm.~4]{manlove_journal}, Thm.~\ref{thm:nonunit_fc_Reduction}
&
P~\cite[Thm.~2]{manlove_journal}, Thm.~\ref{thm:occ_stable_existence}
\\

Max
&
NP-h~\cite[Thm.~3.8]{hrs_hardness}
&
NP-h~\cite[Thm.~4]{manlove_journal}, Thm.~\ref{thm:nonunit_fc_Reduction}
&
NP-h~\cite[Thm.~8]{manlove_journal}, Thm.~\ref{thm:occ_stable_hardness}
\\

ML(H)-Find
&
P~\cite[Prop.~2]{manlove_journal}, 
&
P~\cite[Prop.~2]{manlove_journal}
&
P~\cite[Prop.~2]{manlove_journal}
\\

ML(H)-Max
&
P~\cite[Obs.~3]{manlove_journal}
&
NP-h~\cite[Thm.~7]{manlove_journal}
&
NP-h~\cite[Thm.~8]{manlove_journal}, Thm.~\ref{thm:occ_stable_hardness}
\\

GenML(A)-Find
&
P, Thm.~\ref{thm:gen_master_list}
&
P, Thm.~\ref{thm:gen_master_list}
&
P, Thm.~\ref{thm:gen_master_list}
\\

GenML(A)-Max
&
P, Thm.~\ref{thm:max_size_gen_master_list}
&
NP-h~\cite[Thm.~7]{manlove_journal}
&
NP-h~\cite[Thm.~8]{manlove_journal}, Thm.~\ref{thm:occ_stable_hardness}
\\

\bottomrule
\end{tabularx}
\caption{Summary of known complexity results for \HRS\ across three stability notions (columns) and problem restrictions (rows), together with the new results established in this paper. Here, ML(H) denotes the restriction in which there is a master list defined on hospitals~\cite{manlove_journal}, while GenML(A) denotes the generalised master list defined on agents (introduced in this paper). Row labels ending in \textbf{-Find} denote deciding whether a (variant of) stable matching exists, while those ending in \textbf{-Max} denote finding a maximum-size (variant of) stable matching.}
\label{tab:complexity}
\end{table}

We first consider the decision problem. Deciding whether a first-coalition stable matching exists is \NP-hard~\cite[Thm.~4]{manlove_journal}. In this paper, we establish the same hardness independently (Theorem~\ref{thm:nonunit_fc_Reduction}) under a different restriction, namely that every non-unit sized agent has degree one; neither hardness result implies the other. Since every classically stable matching is first-coalition stable (Proposition~\ref{prop:sfcpsm}), the \NP-hardness of deciding the existence of a classically stable matching follows immediately, recovering the known result~\cite{hrs_hardness}. In contrast, every \HRS\ instance admits an occupancy-stable matching, which can be computed in polynomial time (Theorem~\ref{thm:occ_stable_existence}). An alternate algorithm for computing the same was proposed in~\cite{manlove_journal}. In fact, both algorithms are very similar, except that Algorithm~\ref{algo:gener_ML} is agent-proposing, whereas Algorithm~1 in \cite{manlove_journal} is student-proposing, corresponding to a hospital-proposing algorithm under the \HRS\ correspondence.

We now consider the optimisation problem. Computing a maximum-size classically stable matching is \NP-hard~\cite{hrs_hardness}. This follows immediately from the \NP-hardness of deciding the existence of a classically stable matching. Likewise, computing a maximum-size first-coalition stable matching is \NP-hard (\cite[Thm.~4]{manlove_journal}, and Theorem~\ref{thm:nonunit_fc_Reduction}). Unlike the previous two notions, every \HRS\ instance admits an occupancy-stable matching. We show that computing a maximum-size occupancy-stable matching is \NP-hard even when the preference lists of both agents and hospitals follow a master list via a reduction from \textsc{Vertex Cover} (Theorem~\ref{thm:occ_stable_hardness}). Whereas \cite[Thm.~8]{manlove_journal} establishes the corresponding hardness result via a reduction from \textsc{Common Stable Marriage with Ties and Incomplete Lists} (COM-SMTI).

Master lists, studied in \cite{manlove_journal}, are a common restriction considered in the literature and are motivated by real-world applications. As noted by Irving~\textit{et al.}~\cite{master_list}, the Medical Training Application Service (MTAS) employed score-derived master lists with extensive ties to allocate medical posts. Allocation of students to dormitories at the Technion--Israel Institute of Technology~\cite{ML_application1} makes use of master lists on students. Rodr\'{i}guez and Manlove~\cite{manlove_journal} study the restriction in which hospitals follow a master list, whereas we consider the setting in which agents follow a generalised master list, with a master list on agents as a special case. Under the respective restrictions, both algorithms compute a stable matching and hence also a first-coalition stable and occupancy-stable matching (Prop.~\ref{prop:sfcpsm}). For the optimisation problem, \cite{manlove_journal} showed that the classically stable matching is unique under a master list on hospitals~\cite[Obs.~3]{manlove_journal}. We extend this tractability by proving a Rural Hospitals Theorem for instances admitting a generalised master list on agents (Lemma~\ref{lem:rht_for_genML}), implying that all stable matchings have the same size. Consequently, a maximum-size classically stable matching can be computed in polynomial time (Theorem~\ref{thm:max_size_gen_master_list}).

For first-coalition stability, computing a maximum-size matching is \NP-hard even when the instance has a master list on both hospitals and agents~\cite[Thm.~7]{manlove_journal}. For occupancy stability, we establish \NP-hardness even when both agents and hospitals admit master lists (Theorem~\ref{thm:occ_stable_hardness} and ~\cite[Thm.~8]{manlove_journal}), as discussed earlier. Since a master list is a special case of a generalised master list, these hardness results immediately extend to the generalised master list setting.

\section{Our algorithm} \label{sec:algorithmic_results}

Let $G$ be an \HRS\ instance and let $\{\mathcal{A}_{t_1}, \ldots, \mathcal{A}_{t_f}\}$ be a partition of $\mathcal{A}$ of $G$ such that all agents in set $\mathcal{A}_{t_i}$ have the same size. We denote such a partition of agents as a {\em same-sized} partition. Note that all agents in a given partition have the same size, however, it is not necessary that all agents of the same size belong to a single partition. We present an algorithm that takes as input such an instance $G$ and an ordering on a same-sized partition of $G$ and computes a matching. Later, we show the following two independent results: 
\begin{itemize}
\item In Section~\ref{sec:occ_stable}, we show that a same-sized partition of $\mathcal{A}$ and an ordering on the sets in that partition exists
such that Algorithm~\ref{algo:gener_ML} computes an occupancy-stable matching. We further show that this matching has an approximation guarantee of {strictly better than $3$} with respect to the max-size occupancy-stable matching (Theorem~\ref{thm:occ_stable_existence} and Theorem~\ref{thm:occ_stable_approx}).
\item In Section~\ref{sec:genML} we assume that $G$ admits a generalised
master list ordering on agents. That is, there exists a same-sized partition of agents and an ordering on the sets of this partition. With this ordering as input, Algorithm~\ref{algo:gener_ML} computes a stable matching (Theorem~\ref{thm:gen_master_list}).
\end{itemize}
    
Our algorithm iterates over these sets in the given ordering. For a fixed set $\mathcal{A}_{t_k}$, it constructs an induced subgraph $G_k$ of agents in the set $\mathcal{A}_{t_k}$ and all hospitals. We note that the graph $G_k$ enjoys the property that all agents in the set $\mathcal{A}_{t_k}$ have the same size. Suppose the size of agents in $G_k$ is $s$. The standard Gale and Shapley algorithm~\cite{gale_shapley} that works with unit-sized agents can be adapted to work for uniform sized agents as follows. Agents apply in order of their preference and hospitals accept/reject based on preference. Each time an agent of size $s$ is matched to $h$, the hospital $h$ subtracts $s$ from its available capacity instead of subtracting $1$. When an agent is rejected by a hospital to accept a better preferred agent, there is no change in available capacity since all agents are of the same size. We execute this adapted version of the Gale and Shapley algorithm on $G_k$ to compute a stable matching $M_k$ in $G_k$. The edges in matching $M_k$ computed in the $k$-th iteration are added to the cumulative union of stable matchings computed till then. That is, $M^{(t_k)} = \displaystyle \cup_{r=1}^{k}{M_r}$. It finally returns the result, that is, $M^{(t_f)}$. Algorithm~\ref{algo:gener_ML} gives the pseudo-code. 

\begin{algorithm}

\begin{algorithmic}[1]

\Statex {\bf Input:} an \HRS\ instance $G$ with an ordering $\langle \mathcal{A}_{t_1}, \ldots, \mathcal{A}_{t_f}\rangle$ on the sets in same-sized partition of $\mathcal{A}$

\State $M^{(t_0)} = \emptyset$

\For {$k = 1, \dots, f$}\label{line:for}

\For {$h \in \mathcal{H}$}

\State $q_k(h) \gets q(h) - O_{M^{(t_{k-1})}}(h)$
\EndFor

\State let $G_k$ denote the sub-graph of $G$ induced by the vertices in $\mathcal{A}_{t_k} \cup \mathcal{H}$

\State let $M_k$ be a stable matching obtained in $G_k$ with capacities $q_k$ by executing the adapted version of the Gale and Shapley agent-proposing algorithm 
\label{step:computeSM}

\State $M^{(t_k)} \gets M^{(t_{k-1})} \cup M_k$
\label{line:union}

\EndFor

\State return $M^{(t_f)}$
\end{algorithmic}

\caption{Algorithm for \HRS\ with an ordered partition of agents } 

\label{algo:gener_ML}
\end{algorithm}

Let $M = M^{(t_f)}$ denote the output of Algorithm~\ref{algo:gener_ML}. Before we proceed, we state important observations about our algorithm in the propositions below.

\begin{proposition}\label{prop:disjoint}
    The subgraphs $G_k$ are pairwise edge-disjoint.
\end{proposition}

\begin{proposition}\label{prop:match_a_h}
    An edge $(a,h) \in M$ if and only if $(a, h) \in M_k$ such that $a \in \mathcal{A}_{t_k}$.
\end{proposition}

\begin{proposition}\label{prop:O_h_mono_incr}
    For every hospital $h$, the occupancy of $h$ monotonically increases during the execution of the algorithm, that is, if  $i \leq j$ then $O_{M^{(t_i)}}(h) \leq O_{M^{(t_j)}}(h)$.
\end{proposition}

\noindent A stable matching in the subgraph $G_k$ can be computed in $O(|E_k|)$ time using the adapted Gale and Shapley algorithm described above. Here $E_k$ denotes the set of edges in the graph $G_k$. Since the graphs in each iteration are pairwise edge-disjoint (Proposition~\ref{prop:disjoint}), we conclude that the algorithm runs in
linear time in the size of $G$. 

The agent-proposing adaptation of the Gale and Shapley algorithm used in line~\ref{step:computeSM} of Algorithm~\ref{algo:gener_ML} is for the sake of concreteness. We remark that in line~\ref{step:computeSM} we can use any stable matching in the graph $G_k$ and the proof of correctness in the following section will go through. 

\subsection{Occupancy-stable matchings}\label{sec:occ_stable}

Let $G$ be the given \HRS\ instance. Let $s_1, s_2, \ldots, s_k$ denote the distinct sizes of agents in $G$ and let $s_1 > s_2 > \ldots > s_k \geq 1$. Let $\mathcal{A}_{s_i}$ denote the set of agents with size $s_i$. It is clear that $\{\mathcal{A}_{s_1}, \ldots, \mathcal{A}_{s_k}\}$ is a {same-sized} partition of $\mathcal{A}$. Consider the ordering $\langle\mathcal{A}_{s_1}, \ldots, \mathcal{A}_{s_k}\rangle$ on the above partition. We note that such a same-sized partition and an ordering can be computed in $O(n \log{n})$ time where $n = |\mathcal{A} \cup \mathcal{H}|$. We execute Algorithm~\ref{algo:gener_ML} using this ordering imposed in the input instance and show that Algorithm~\ref{algo:gener_ML} computes an occupancy-stable matching in this case. We note that the partition is only with respect to the sizes and here we do  {\em not} have any restriction on the preferences of the hospitals with respect to this partition. Recall that we let $M = M^{(t_f)}$ denote the output of Algorithm~\ref{algo:gener_ML}.

\begin{lemma}\label{lem:weak_stable}
    The matching $M$ computed by Algorithm~\ref{algo:gener_ML} is occupancy-stable.
\end{lemma}

\begin{proof}
    Suppose, for the sake of contradiction, that $(a,h)$ is an occupancy-blocking pair w.r.t. $M$. This implies that $h \succ_a M(a)$ and one of the following two conditions must hold: 
    \begin{itemize}
        \item (i) $O_M(h) + s(a) \leq q(h)$ or 
        \item (ii) there exists a non-empty set of agents $X \subseteq M(h)$ such that for every $a' \in X$, $a \succ_h a'$ and $s(a) \geq \displaystyle \sum_{a' \in X}{s(a')}$ and $O_M(h) - \displaystyle \sum_{a'\in X}{s(a')} + s(a) \leq q(h)$.
    \end{itemize}
    
    We show that neither of (i) and (ii) holds, thereby leading to a contradiction
    to the claimed occupancy-blocking pair.

    Let $s_a = s(a)$. Since $(a,h)$ is an occupancy-blocking pair, $(a,h) \notin M$. Therefore, by Proposition~\ref{prop:match_a_h}, $(a,h) \notin M_{s_a}$. Consider the iteration when $k = s_a$. Recall that $O_{M^{(t_k)}}(h)$ denotes the occupancy of $h$ in matching $M^{(t_k)}$, that is, $O_{M^{(t_k)}}(h)$ denotes the occupancy of $h$ after computing the union in line~\ref{line:union}. Since $(a,h) \notin M_{s_a}$, we have $O_{M^{(t_k)}}(h) + s(a) > q(h)$. By Proposition~\ref{prop:O_h_mono_incr}, $O_M(h) = O_{M^{(t_f)}}(h) \geq O_{M^{(t_k)}}(h)$, thus, $O_M(h) + s(a) > q(h)$ holds at the end of the algorithm. Therefore, (i) does not hold at the end of the algorithm.

    Next, we show that (ii) does not hold either. We first consider the case wherein for every agent $a' \in X$, $s(a') = s_a$. Due to the stability of matching $M_{s_a}$, for all such agents, $a' \succ_h a$ holds. This leads to a contradiction to the given condition that for every $a' \in X$, $a \succ_h a'$. Therefore, there must exist an agent $a' \in X$ such that $s(a') \neq s(a)$. 
    
    First, suppose that there exists an agent $b \in X$ such that $s(b) > s(a)$. Then, clearly, $\displaystyle \sum_{a' \in X}{s(a')} \geq s(b) > s(a)$. This leads to a contradiction to the given condition that $s(a) \geq \displaystyle \sum_{a' \in X}{s(a')}$. Therefore, for every $a' \in X$, $s(a') < s(a)$. This implies that all such agents $a'$ are matched to $h$ {\em after} $M_{s_a}$ and $M^{(t_k)}$ were computed. 
        
    Since $X \subseteq M$, we have $\displaystyle O_{M^{(t_k)}}(h) + \sum_{a' \in X}{s(a')} \leq O_M(h)$. Therefore, $\displaystyle O_M(h) - \sum_{a'\in X}{s(a')} \geq O_{M^{(t_k)}}(h)$. Adding $s(a)$ to both sides, we get $\displaystyle O_M(h) - \sum_{a'\in X}{s(a')} + s(a) \geq O_{M^{(t_k)}}(h) + s(a)$. As observed earlier in this proof, $O_{M^{(t_k)}}(h) + s(a) > q(h)$, therefore, $\displaystyle O_M(h) - \sum_{a'\in X}{s(a')} + s(a) > q(h)$. This leads to a contradiction to the given condition that $O_M(h) - \displaystyle \sum_{a'\in X}{s(a')} + s(a) \leq q(h)$. Therefore, (ii) does not hold. This completes the proof.
\end{proof}

Before proving the approximation guarantee, we illustrate the execution of Algorithm~\ref{algo:gener_ML} on an example. Consider the \HRS\ instance shown in Fig.~\ref{fig:ex_2bound}. For this instance, $\mathcal{A}$ is partitioned as $\{\mathcal{A}_{t_1}, \mathcal{A}_{t_2}\}$ such that $\mathcal{A}_{t_1} = \{a_1\}, \mathcal{A}_{t_2} = \{a_2, a_3\}$ where $\mathcal{A}_{t_1}$ contains all agents of size $3$ and $\mathcal{A}_{t_2}$ contains all agents of size $2$. Algorithm~\ref{algo:gener_ML} computes $M_1 = \{(a_1,h_1)\}$ and $M_2 = \emptyset$, thereby giving the output $M = M^{(t_2)} = \{(a_1,h_1)\}$. The matching $M^* = \{(a_1, h_2), (a_2, h_1), (a_3, h_1)\}$ is the maximum size occupancy-stable matching in this instance. Thus, for this instance, we get an approximation ratio of $\frac{s(M^*)}{s(M)} = \frac{7}{3} < 3$. In what follows we prove that Algorithm~\ref{algo:gener_ML} gives an approximation guarantee of $2+\rmax$ (defined below) where $\rmax < 1$.
\begin{figure}[!ht]
\begin{align*}
    (3) \ a_1 &: h_1 \succ h_2 & (2) \ a_2 &: h_1 & (2) \ a_3 &: h_1 \\
    [4] \ h_1 &: a_2 \succ a_3 \succ a_1 & [3] \ h_2 &: a_1
\end{align*}
\caption{Example of \HRS\ instance used to illustrate the execution of Algorithm~\ref{algo:gener_ML}.}
\label{fig:ex_2bound}
\end{figure}

\noindent \textbf{Approximation guarantee.} 
Recall that the distinct sizes in the instance are denoted as $s_1 > s_2 > \ldots > s_k \geq 1$.
In conference version~\cite{balasundaram_et_al_fsttcs_2025} of our work, we proved an approximation guarantee of
$3-\frac{1}{\smax}$, that is, $3-\frac{1}{s_1}$. For the instance in Fig.~\ref{fig:ex_2bound}, this guarantee is $3-\frac{1}{3} = \frac{8}{3}$ whereas, as seen earlier, the achieved ratio is $\frac{7}{3}$. In this work, we improve the analysis which is tight for the example in Fig~\ref{fig:ex_2bound}.
We define $\rmax$ as below:
\begin{eqnarray*}
\rmax = \max\limits_{2\leq j \leq k} \frac{s_{j}-1}{s_{j-1}}
\end{eqnarray*}

For a fixed size $s$, the ratio is simply the $s-1$ divided by the next largest size. We consider every size $s$ starting with the second largest size and pick the ratio that achieves the maximum value. This is $\rmax$ and it is clear that $\rmax < 1$. 
For the example in Fig.~\ref{fig:ex_2bound}, $\rmax = \frac{2-1}{3} = \frac{1}{3}$, thus the approximation guarantee is $2+\rmax = \frac{7}{3}$. This guarantee 
matches the achieved ratio of $\frac{7}{3}$. 

We prove this result in general, that is, the matching $M$ computed by Algorithm~\ref{algo:gener_ML} is a $(2+\rmax)$-approximation for the maximum size occupancy-stable matching. 
Moreover, when the instance has sizes $s_1 = s > 1, s_2 = 1$, we get $\rmax = 0$, hence the approximation guarantee is $2$. We note that the guarantee of $2$ is independent of the value of $s$, the largest size in such instances.

Recall that the size of a matching is the total occupancy across all hospitals. Let $s(N)$ denote the size of matching $N$. Then $s(N) = \displaystyle \sum_{h \in \mathcal{H}}{O_N(h)} = \sum_{a \in \mathcal{A}, N(a) \neq \bot}{s(a)}$.

\begin{lemma}\label{lem:3approx}
    Let $M^*$ be a maximum size occupancy-stable matching in $G$. Then $s(M^*) \le s(M) \cdot (2+\rmax)$.
\end{lemma}

\begin{proof}
We partition the set of agents matched in $M^*$ as follows: 
\begin{itemize}
    \item $X = \{a \mid \mbox{ $a$ is matched in both $M$ and $M^*$} \}$
    \item $Y = \{a \mid  \mbox{ $a$ is matched in $M^*$ but $a$ is not matched in $M$} \}$
\end{itemize}

Clearly, $s(M^*) = \displaystyle \sum_{a \in X}{s(a)} + \sum_{a \in Y}{s(a)}$. We note that $\displaystyle \sum_{a \in X}{s(a)} \leq s(M)$, therefore $s(M^*) \leq s(M) + \displaystyle \sum_{a \in Y}{s(a)}$. In Claim~\ref{cl:boundsY}, we show that $\sum_{a \in Y}{s(a)} \leq s(M) (1+\rmax)$, thereby proving the lemma.
\end{proof}

\begin{claim}\label{cl:boundsY}
Let $s(Y) = \sum_{a \in Y}{s(a)}$. Then, $s(Y) \leq s(M)(1+\rmax)$.
\end{claim}

Before proving Claim~\ref{cl:boundsY}, we prove two crucial properties (Claim~\ref{cl:noworse} and Claim~\ref{cl:M_guarantee}) of the computed matching $M$.
Recall that $s_1$ denotes the size of the largest sized agent in the instance. 
We first establish via Claim~\ref{cl:noworse} that for any agent of size $s_1$ the partner that $a$ is assigned to in our matching $M$ is at least good as the partner that $a$ is assigned to {\em any} occupancy stable matching $M'$.

\begin{claim}
\label{cl:noworse}
Let $M'$ be any occupancy stable matching and $M$ be the output of our algorithm. Let $a$ be any agent of the largest size, that is, $s(a) = s_1$. Then, $M(a) \succeq_a M'(a)$.  
\end{claim}

\begin{proof}
Let $A'$ denote a set of agents of size $s_1$ such that for every $a \in A'$ we have $M'(a) \succ_a M(a)$. For the sake of contradiction, assume that the set $A'$ is non-empty. During the course of our algorithm when the Gale and Shapley round was executed for size $s_1$, each $a' \in A'$ must have proposed to the respective $M'(a')$ and was rejected. Amongst all such rejections during our execution sequence, consider the first such rejection of $a'$ by say  $h' = M'(a')$. The rejection of $a'$ by $h'$ was for a better preferred agent $\tilde{a}$ where $\tilde{a} \succ_{h'} a$.  Note that since it is the first round in our algorithm, that is, where largest size agents are considered, we know that $s(\tilde{a}) = s_1$. We now remark that $M'(\tilde{a}) \succ_{\tilde{a}} M(\tilde{a})$ , otherwise, $(\tilde{a}, h')$ is an occupancy blocking pair for $M'$. 
However, in our algorithm the proposal by $\tilde{a}$ to $M'(\tilde{a})$ must have happened before $\tilde{a}$ proposed to $h'$. Moreover, $M'(\tilde{a})$ must have rejected $\tilde{a}$ since $\tilde{a}$ proposed to $h'$. This contradicts that the rejection of $a'$ by $M'(a')$ was the first such rejection by our algorithm.
This completes the proof.
\end{proof}
By the above claim the corollary below is immediate.
\begin{corollary}\label{corr:smax}
Let $a$ be an agent of size $s_1$ that is left unmatched in our matching $M$. Then agent $a$ cannot be matched in any occupancy stable matching.
\end{corollary}

 For a fixed a size $s$, a hospital $h$, we define a set $A_s^*(h)$ of agents all of them having size $s$  that are matched to $h$ in the optimal matching $M^*$ and all these agents prefer $h$ over their matched partner in our matching $M$. In particular, these agents were rejected by $h$ during the execution of  Algorithm~\ref{algo:gener_ML} in Step~\ref{step:computeSM}.
In the claim below we establish a useful property about the matched partners of $h$ in our matching $M$.  
\begin{claim}\label{cl:M_guarantee}
Let $s$ be any fixed size and $h$ be any hospital. Let $M^*$ be an optimal occupancy stable matching. Define
\begin{eqnarray*}
A^*_{s}(h) = \{ a \mid a \in M^*(h) \mbox { and $a$ prefers $h$ over $M(a)$ and $a$ has size $s$} \} 
\end{eqnarray*}
Let $t_{s}(h) = |A^*_{s}(h)|$. 
If $t_s(h) > 0$ then one of the following holds:
\begin{enumerate}
    \item \label{cond:one} $M(h)$ contains at least one agent $c$ such that $s(c) > s$.
    \item \label{cond:two} $M(h)$ contains at least $t_s(h)$ many agents of size $s$ such that $h$ prefers all of them over all agents in $A^*_s(h)$.
\end{enumerate}
\end{claim}
\begin{proof}
    If $M(h)$ contains at least one agent $c$ such that $s(c) > s$, then we are done.
    Otherwise, suppose $s$ is the $k'$-th largest size in the instance, that is, $s = s_{k'}$. By assumption, no agent was matched to $h$ before the iteration of the {\bf for} loop in line~\ref{line:for} with $k=k'$ began. All agents matched to $h$ in $M_{k'}$ have size $s$ each. Moreover, owing to stability of $M_{k'}$ and the fact that $h$ rejected all agents in $A^*_s(h)$ , we conclude that agents matched to $h$ in $M_{k'}$ are higher preferred by $h$ over all agents in $A^*_s(h)$. It remains to show that the number of agents in $M_{k'}$ is at least $t_s(h)$.
    
    Clearly, $q(h) \geq O_{M^*}(h)$. Since $A^*_s(h) \subseteq M^*(h)$, we have $q(h) \geq O_{M^*}(h) \geq |A^*_s(h)| \cdot s = t_s(h) \ \cdot \ s$. Since we assumed that $M(h)$ does not contain any agent of size larger than $s$,  the entire capacity of $q(h)$ was {\em available} during the $k=k'$ iteration of the Gale and Shapely algorithm. 
    Thus, $q_{k'}(h) = q(h)$. Therefore, in $M_{k'}$, at least $t_s(h)$ many agents are matched to $h$. Otherwise $h$ has enough capacity to accommodate at least one more agent of size $s$, thereby forming a blocking pair with every agent in $A^*_s(h)$ hence contradicting the stability of $M_{k'}$.
\end{proof}
For a fixed size $s$ and a hospital $h$ such that $|A^*_s(h)| = t_s(h) > 0$, and Condition~\ref{cond:one} of Claim~\ref{cl:M_guarantee} does not apply, we construct a one-to-one mapping of agents in $A^*_s(h)$ to $s$-sized agents in $M(h)$. We define this function as $map_{s,h} (a) = a'$ where $a \in A^*_s(h), a' \in M(h), s(a) = s(a') = s$ and the hospital $h$ prefers $a'$ over $a$, that is, $a' \succ_h a$.

Now we proceed to prove Claim~\ref{cl:boundsY}.
Before we present a formal proof, we give a sketch below:
we partition the set $Y$ as $\{Y_1, Y_2, \ldots, Y_t\}$ such that
each $Y_j \neq \emptyset$. Let $s(Y_j) = \sum_{b\in Y_j}{s(b)}$.
Then, to upper-bound the $s(Y)$, we upper bound $\sum_{j=1}^{t}{s(Y_j)}$.
We obtain the partition of $Y$ as follows: starting with an agent $b \in Y$, we construct
edge-disjoint alternating paths. 
Each path starts at an agent $b \in Y$ and terminates at a hospital $h_j$ and all agents along the path have the same size $s(b)$, that is, the size of agent $b$. Moreover, the path starts and ends with an $M^*$-edge and the terminating hospital $h_j$ is such that $M(h_j)$ contains an agent whose size is strictly larger than $s(b)$.

The set $Y_j$ in the partition is simply the set of all agents whose paths terminate at the fixed hospital $h_j$. The structure of these paths allow us to upper bound $s(Y_j)$ by $q(h_j)$,
which is further upper bounded by $O_M(h_j)\cdot (1+\rmax)$. Noticing that $s(M) = \sum_h{O_M(h)}$ and combining these results, we prove the required upper bound on $s(Y)$ as $s(M)(1+\rmax)$.

\begin{proof}[Proof of Claim~\ref{cl:boundsY}]

{
    For the sake of analysis, we construct edge-disjoint alternating paths starting at every agent in $Y$.
    {See Fig.~\ref{fig:alt_path} for an illustration.}
    Let $b\in Y$ and $s = s(b)$. Let $a_0 = b$ and $h_0 = M^*(a_0)$. 
    By the definition of $Y$, $h_0$ exists.
    We start the path $P$ with an $M^*$-edge $(a_0, h_0)$.
    We note that since $a_0$ is unmatched in $M$,
    $h_0 \succ_{a_0} M(a_0)$. 
    Thus, by the definition of $A^*_s(h_0)$, we have that $a_0 \in A^*_s(h_0)$.
    If there exists an agent $c \in M(h_0)$ such that $s(c) > s$, 
    we terminate the alternating path $P$ at $h_0$ itself.}
    {Otherwise, condition~\ref{cond:two} of Claim~\ref{cl:M_guarantee} must hold.
    Therefore, $map_{s,h_0}(a_0)$ is well-defined.}
    
    {Let $a_1 = map_{s,h_0}(a_0)$.
    By definition of $a_1$, we have that $a_1 \in M(h_0), s(a_1) = s(a_0)$
    and $a_1 \succ_{h_0} a_0$.
    {We extend the path $P$ by adding the edge $(h_0, a_1)$.}
    Thus, so far we have constructed an alternating path $P = \langle a_0, h_0, a_1 \rangle$
    where $(a_0, h_0) \in M^*$ and $(h_0, a_1) \in M$.
    Since $s(a_1) = s(a_0)$ and $a_1 \succ_{h_0} a_0$, occupancy-based stability of $M^*$ implies that
    $a_1$ is matched in $M^*$ to a higher preferred hospital than $h_0$. Let $h_1 = M^*(a_1)$.
    We extend the alternating path $P$ by adding the $M^*$-edge $(a_1, h_1)$ and repeat the process at $a_1$.}
    {We note that $a_1 \notin Y$, but $a_1$ prefers its $M^*$-partner over its $M$-partner. Thus, $a_1 \in A^*_s(h_1)$, hence Claim~\ref{cl:M_guarantee} applies to $a_1$.}

    That is, for a fixed agent $a_0 \in Y$ of size $s$, we construct an alternating path of the form $\langle a_0, h_0, a_1, h_1, a_2, h_2 \ldots h_j \rangle$ where $(a_i, h_i) \in M^*$ and $(a_{i+1}, h_i) \in M$ and all agents in this path have size equal to $s(a_0)$.
    {Next, we argue that the alternating paths constructed this way are edge-disjoint: }
    {while constructing the alternating paths, if we are at an agent $a_\ell$ then there is a fixed $M^*$-edge 
    $(a_\ell, h_\ell = M^*(a_\ell))$ by which the path gets extended. Whereas, if we are at a hospital $h_\ell$ then the mapping $map_{s,h_\ell}(a_\ell)$ gives the distinct agent $a_{\ell+1}$ such that the path gets extended by the $M$-edge $(h_\ell, a_{\ell+1})$.
    Therefore, the paths constructed are edge-disjoint.
    Moreover, since there are finitely many agents for a fixed size $s$, every alternating path {starting at an agent with size $s$} must terminate. Moreover, such a path must terminate at a hospital $h_j$ such that there exists an agent $c \in M(h_j)$ such that $s(c) > s$.}
    {We note here that the path terminates at $h_j$ and the edge $(h_j, c)$ is {\em not} included in the path.}

 Once alternating paths from every $b \in Y$ are constructed, we partition the set $Y$ by grouping agents
    w.r.t. the hospital $h_j$ at which their alternating paths terminated.
    {That is, we let $Y_j \subseteq Y$ to be the set of all agents $b \in Y$ whose alternating path terminates at the fixed hospital $h_j$.}
    By definition of $Y_j$, we know that $Y_j \neq \emptyset$.
    {Recall that $s(Y_j) = \sum_{b\in Y_j}{s(b)}$.}
    Assume that $Y = Y_1 \cup Y_2 \cup \ldots \cup Y_t$.
    Our goal is to show that for each $Y_j$ in the partition of $Y$, $s(Y_j) \leq O_M(h_j)\cdot (1+\rmax)$.
    Once we show that, we prove the claim as follows:
    \begin{align*}
        s(Y) &= \sum_{j=1}^{t}{s(Y_j)} \leq \sum_{j=1}^{t}{O_M(h_j) \cdot (1+\rmax)} \leq \left(1+\rmax\right) \sum_{j=1}^{t}{O_M(h_j)} \\
        &\leq \left(1+\rmax\right) \cdot \sum_{h}{O_M(h)} = \left(1+\rmax\right) \cdot s(M)
    \end{align*}
    Now we proceed to show that $s(Y_j) \leq O_M(h_j)\cdot (1+\rmax)$.
    {See Fig.~\ref{fig:grouping} for an illustration of alternating paths starting at agents in $Y_j$ all of which terminate at $h_j$.} Suppose that the alternating paths for agents $b_1, \ldots, b_r$ terminate at hospital $h_j$.
    Then $Y_j = \{b_1, \ldots, b_r\}$.
    {In Fig.~\ref{fig:grouping}, $Y_j = \{b_1, b_2, b_3\}$.}
    Let $d_i$ denote the last agent occurring on the path starting from agent $b_i$.
    Then, we have that $M^*(d_i) = h_j$.
    We note that, along each alternating path, the size of all agents is the same as the size of $b_i$ with which the path started. Hence, $s(d_i) = s(b_i)$ on the fixed path starting at $b_i$. 
    However, the sizes of $b_1, \ldots, b_r$ need not be the same.

    Since the paths are edge-disjoint, we have that $s(Y_j) = \sum\limits_{b_i\in Y_j}{s(b_i)} = \sum\limits_{d_i}{s(d_i)}$.
    Further, since $(d_i, h_j)$ edges are matched in $M^*$, we have $\sum\limits_{d_i}{s(d_i)} \leq O_{M^*}(h_j)$.
    Thus, $s(Y_j) \leq O_{M^*}(h_j)$. Observing that $O_{M^*}(h_j) \leq q(h_j)$, we have that
    $s(Y_j) \leq q(h_j)$. 

\begin{minipage}{0.45\textwidth}
    \begin{figure}[H]
       \centering
       \begin{tikzpicture}
            \node (a0) at (0,0) {$b=a_0$};
            \node (h0) at (1,1.1) {$h_0$};
            \node (a1) at (0,2.2) {$a_1$};
            \node (h1) at (1,3.3) {$h_1$};
            \node (dummy1) at (0.5,3.6) {};
            \node (dummy2) at (0.5,4.2) {};
            \draw[thick,loosely dashed] (a0) -- node[above,pos=0.01] {$+$} (h0);
            \draw[thick] (h0) -- (a1);
            \draw[thick,loosely dashed] (a1) -- node[above,pos=0.01] {$+$} (h1);
            \draw[thick,dotted] (dummy1) -- (dummy2);
       \end{tikzpicture}
       \caption{Alternating path starting at $b\in Y$, as constructed in the proof of Claim~\ref{cl:boundsY}. The $M^*$ edges are dashed and $M$ edges are solid.
       By the construction of these paths, each agent prefers its $M^*$-partner over its $M$-partner, indicated by $+$ on the dashed edge.}
       \label{fig:alt_path}
   \end{figure}
   \end{minipage}
   \hfill
\begin{minipage}{0.45\textwidth}
    \begin{figure}[H]
       \centering
       \begin{tikzpicture}
            \node (a0) at (0,0) {$a_0=b_1$};
            \node (h0) at (1,1) {$h_0$};
            \node (a1) at (0,2) {$a_1=d_1$};
            \node (hj) at (3,5) {$h_j$};
            \node (a00) at (2,0) {$a_0' = b_2 = d_2$};
            \node (a01) at (4,0) {$a_0'' = b_3$};
            \node (h01) at (5,1) {$h_0''$};
            \node (a11) at (4,2) {$a_1''$};
            \node (a21d3) at (4,4) {$d_3$};
            \draw[thick,loosely dashed] (a0) -- node[above,pos=0.05] {$+$} (h0);
            \draw[thick] (h0) -- (a1);
            \draw[thick,loosely dashed] (a1) -- node[left,pos=0.03] {$+$} (hj);
            \draw[thick,loosely dashed] (a00) -- node[left,pos=0.03] {$+$} (hj);
            \draw[thick,loosely dashed] (a01) -- node[above,pos=0.01] {$+$} (h01);
            \draw[thick] (h01) -- (a11);
            \draw[thick,dotted] (a11) -- (a21d3);
            \draw[thick,loosely dashed] (a21d3) -- node[above,pos=0.01] {$+$} (hj);
       \end{tikzpicture}
       \caption{An illustration of alternating paths starting at agents $b_1, b_2, b_3$ in $Y_j$. The paths are $\langle b_1=a_0, h_0, a_1=d_1, h_j\rangle$,$\langle b_2=a_0'=d_2, h_j\rangle$ and $\langle b_3=a_0'', h_0'', a_1'', \ldots, d_3, h_j\rangle$. Recall that $+$ on the dashed edge indicates that every agent prefers their $M^*$-partner over their $M$-partner. The last agents along these paths, namely $d_1, d_2, d_3$, are shown.}
       \label{fig:grouping}
   \end{figure}
   \bigskip
   \end{minipage}
    
    Since the paths from agents in $Y_j$ terminate at $h_j$, we know that there exists an agent $c \in M(h_j)$ such that $s(c) > \max\limits_{b_i \in Y_j}{s(b^i)}$. Therefore, the following holds:
    \begin{align}\label{eq:sc}
        O_M(h_j) \geq s(c) > \max\limits_{b_i \in Y_j}{s(b^i)} \geq \max\limits_{b_i \in Y_j}{s(b^i)} + 1
    \end{align}
    Further, none of the $d_i$ agents are matched to $h_j$ in $M$ but they prefer $h_j$ over $M(d_i)$. Combining this with the fact that $M$ is occupancy-stable, we know that matching any of the $d_i$ agents to $h_j$ in $M$ violates the capacity of $h_j$.
    That is, $q(h_j) \leq O_M(h_j) + \left(\min\limits_{d_i}{s(d_i)}\right) - 1$. Since $s(d_i) = s(b_i), $
    we have that $q(h_j) \leq O_M(h_j) + \left(\min\limits_{b_i \in Y_j}{s(b_i)}\right) - 1$.
    Let $s^{min} = \min\limits_{b_i \in Y_j}{s(b_i)}$.
    Therefore, we get that

    \begin{align*}
        q(h_j) &\leq O_M(h_j) + s^{min} - 1 = O_M(h_j) \left(1 + \frac{s^{min} - 1}{O_M(h_j)}\right)
                \leq O_M(h_j) \left(1 + \frac{s^{min} - 1}{s(c)}\right)
    \end{align*}

\noindent
Recall from Eq.~\ref{eq:sc} that $s(c) \geq \max\limits_{b_i \in Y_j}{s(b_i)}+1$, and by definition of $s^{min}$, we know that $\max\limits_{b_i \in Y_j}{s(b_i)} \geq s^{min}$, we have $s(c) \geq s^{min}+1$. 
Noting that $s(c)$ is a size in the input instance,
we have that $s(c)$ is at least the immediate next size larger than $s^{min}$ {in the input instance}.
Let's call that size $next(s^{min})$.
Moreover, since $s^{min}$ is the size of some agent in $Y$, by Corollary~\ref{corr:smax}, $s^{min} \leq s_2$.
{Noting that $s_1 > s_2 > s_3 > \ldots > s_k$ and $s^{min} \in \{s_2, s_3, \ldots, s_k\}$}, we get the following:

    \begin{align*}
        q(h_j) \leq O_M(h_j) \left(1 + \frac{s^{min} - 1}{next(s^{min})}\right) &= O_M(h_j) \left(1 + \max\limits_{2\leq i\leq k}\frac{s_i - 1}{next(s_i)}\right) \\
        &= O_M(h_j) \left(1 + \max\limits_{2\leq i\leq k}\frac{s_i - 1}{s_{i-1}}\right) \\
        &= O_M(h_j)\ (1+\rmax)
    \end{align*}

We get the second equality by observing that the maximum ratio is achieved for some size $s_i$ in the instance where $2 \leq i \leq k$.
As observed earlier, $s(Y_j) \leq q(h_j)$. Therefore, $s(Y_j) \leq O_M(h_j) \left(1+\rmax\right)$.
This completes the proof.
\end{proof}

Recall that computing an ordering on the same-sized partition takes $O(n \log{n})$ time, and Algorithm~\ref{algo:gener_ML} runs in $O(m + n)$ time. These facts along with Lemma~\ref{lem:weak_stable} establish Theorem~\ref{thm:occ_stable_existence}. Moreover, Lemma~\ref{lem:3approx} establishes Theorem~\ref{thm:occ_stable_approx}.

\subsection{Stable matchings under generalised master list setting}\label{sec:genML}

In this section, we consider \HRS\ instances that admit a generalised master list ordering on agents. This implies that there exists a same-sized partition $\langle\mathcal{A}_{t_1}, \mathcal{A}_{t_2}, \mathcal{A}_{t_3}, \ldots\rangle$ of $\mathcal{A}$ and an ordering on it such that every hospital $h \in \mathcal{H}$ prefers its neighbours in $\mathcal{A}_{t_i}$ over its neighbours in $\mathcal{A}_{t_j}$ iff $i < j$. We show that Algorithm~\ref{algo:gener_ML} outputs a stable matching when such an ordering is given as input.

\begin{lemma}\label{lem:gen_ML_stable}
    Suppose that $G$ admits a generalised master list ordering on agents. When this ordering is given as input, the matching $M$ computed by Algorithm~\ref{algo:gener_ML} is stable.
\end{lemma}

\begin{proof}
Assume for the sake of contradiction that there is a blocking pair $(a, h)$ w.r.t. the matching $M$. Suppose that $a \in \mathcal{A}_{t_i}$ for some $i$. Since $(a,h)$ blocks $M$, $(a,h) \notin M$. By Proposition~\ref{prop:match_a_h}, $(a,h) \notin M_{t_i}$ that was computed during the iteration when $k = i$.

Since $(a,h)$ blocks $M$, there exists a set $X \subseteq M(h)$ such that for every $a' \in X$, $a \succ_h a'$ and $O_M(h) - \sum_{a' \in X}{s(a')} + s(a) \leq q(h)$. Let $s(X) = \sum_{a' \in X}{s(a')}$. Since $(a,h) \notin M_{t_i}$, we have $O_{M^{(t_i)}}(h) + s(a) > q(h)$. Moreover, due to the stability of $M_{t_i}$ and due to the assumption that the input admits a generalised master list ordering on agents, $h$ prefers all agents matched to it in $M^{(t_i)}$ over $a$. Therefore, all agents in $X$ must be matched to $h$ {\em after} the $i^{th}$ iteration as they are lower-preferred by $h$ over $a$. To complete the proof, we need to show that even if we unmatch all agents in $X$, $h$ cannot accommodate the agent $a$, thereby getting a contradiction to the claimed blocking pair.

By Proposition~\ref{prop:O_h_mono_incr}, $O_M(h) \geq O_{M^{(t_i)}}(h)$. In fact, since all agents in $X$ are matched to $h$ after $M^{(t_i)}$ was computed, we have $O_M(h) \geq O_{M^{(t_i)}}(h) + s(X)$. Thus, $O_M(h) - s(X) \geq O_{M^{(t_i)}}(h)$. Substituting $O_{M^{(t_i)}}(h) > q(h) - s(a)$, we get $O_M(h) - s(X) > q(h) - s(a)$. Hence, $O_M(h) - s(X) + s(a) > q(h)$, leading to contradiction that $O_M(h) - s(X) + s(a) \leq q(h)$.
\end{proof}

We assume that the ordering on the partitions implied by the generalised master list is part of the input instance $G$. This assumption, together with the fact that Algorithm~\ref{algo:gener_ML} runs in $O(m + n)$ time establish Theorem~\ref{thm:gen_master_list}.

\subsubsection{Analogue of Rural Hospitals Theorem for \HRS\ with generalized master lists }
Next, we establish an analogue of the Rural Hospitals Theorem for stable matchings in \HRS\ instances with a generalised master list. As a consequence of our theorem, all stable matchings in an HRS instance with a generalised master list match the {\em same set of agents}. Thus all stable matchings are of the same size and the stable matching computed by Algorithm~\ref{algo:gener_ML} is of maximum size. 

We need to define {\em under-subscription} of hospitals for \HRS\ Setting. Intuitively, a hospital is \emph{under-subscribed} if it can accommodate one of the acceptable agents that is not matched to it already. Formally, for a hospital $h$ and a matching $M$, $s_{M}(h)$ be the smallest size of the agents in $h$'s preference list that are not matched to $h$ in $M$. We say that $h$ is \emph{under-subscribed} with respect to $M$ if $q(h)-O_M(h)\ge s_{M}(h)$.

We also observe a deviation from the classical Rural Hospitals Theorem. In the classical setting an under-subscribed hospital is matched to the same set of agents across all stable matchings. This need not hold as illustrated by the example in Fig.~\ref{fig:rural_hosp_thm}.  

\begin{figure}[H]
\begin{align*}
   (2) \ a_1 &: h_1 \succ h_2 & (2)\  a_2 &: h_2 \succ h_1 &  (1) \ a_3 &: h_3 \succ h_1\\
   [3] \ h_1 &: a_2 \succ a_1 \succ a_3 & [2] \ h_2 &: a_1 \succ a_2 & [1] \ h_3 &: a_3
\end{align*}
\caption{\HRS\ instance with generalised master list $\langle A_1,A_2 \rangle$ where $A_1=\{a_1,a_2\}$ and $A_2=\{a_3\}$. Consider stable matchings $M_1=\{(a_1,h_1),(a_2,h_2),(a_3,h_3)\}$ and $M_2=\{(a_2,h_1),(a_1,h_2),(a_3,h_3)\}$, $h_1$ is under-subscribed by definition but is matched to different agents in two stable matchings.}
\label{fig:rural_hosp_thm}
\end{figure}

Let $I$ be an \HRS\ instance with a generalised master list ordering $\langle A_1,A_2,\ldots,A_f\rangle$, and let $M$ be a stable matching of $I$. For each partition $A_i$, we define the \emph{induced subinstance} of $A_i$ with respect to $M$, denoted by $I_i(M)$. The set of agents in $I_i(M)$ is $A_i$, the set of hospitals is $H$, and the preference lists are obtained by restricting those of $I$. Let $B_i(h)=M(h)\cap\bigcup_{j<i}A_j$ denote the set of agents from higher-preferred partitions that are matched to hospital $h$ in $M$. The capacity of $h$ in $I_i(M)$ is $q_i^M(h)=q(h)-\sum_{a\in B_i(h)}s(a).$ Thus, the capacity of each hospital is reduced by the total size of agents from higher-preferred partitions assigned to it. Finally, let $M_i=M\cap(A_i\times H)$
denote the restriction of $M$ to $I_i(M)$.

\begin{lemma}
\label{lem:restriction_stable}
Let $I$ be an \HRS\ instance with a generalised master list ordering $\langle A_1,A_2,\ldots,A_k\rangle$, and let $M$ be a stable matching of $I$. For every partition $A_i$, the restriction $M_i$ is a stable matching of the induced subinstance $I_i(M)$.
\end{lemma}

\begin{proof}
The matching $M_i$ is feasible in $I_i(M)$ by construction. Now we show that $M_i$ is also stable. Assume, for contradiction, that $M_i$ is not stable in $I_i(M)$. Then there exists a blocking pair $(a,h)$ with respect to $M_i$. We claim that $(a,h)$ is also a blocking pair with respect to $M$ in the original instance $I$, contradicting the stability of $M$.

Since the preference lists in $I_i(M)$ are obtained by restricting those of $I$, agent $a$ prefers $h$ to its assignment in $M$, or is unmatched in $M$. It therefore remains to show that hospital $h$ can accommodate $a$, possibly after removing a set of agents that it prefers less than $a$.

Since $(a,h)$ is a blocking pair in $I_i(M)$, either hospital $h$ is matched in $M_i$ to an agent $a'$ that it prefers less than $a$, or it has sufficient residual capacity to accommodate $a$. In the former case, since all agents in $A_i$ have the same size, unmatching $a'$ creates sufficient capacity to accommodate $a$. In the latter case, $h$ already has sufficient residual capacity to accommodate $a$.

Now consider the original matching $M$. By construction of $I_i(M)$, the capacity occupied by agents from higher-preferred partitions has already been deducted. Therefore, if $h$ can accommodate $a$ in $I_i(M)$ by replacing $a'$, then the same replacement is feasible in $M$. Otherwise, if $h$ can accommodate $a$ directly in $I_i(M)$, then in $M$, if necessary, it can obtain sufficient capacity by removing agents from lower-preferred partitions, all of whom are less preferred than $a$ by the generalised master list ordering. Hence, in either case, hospital $h$ can accommodate $a$ while removing only agents that it prefers less than $a$. Thus, $(a,h)$ is also a blocking pair with respect to $M$, contradicting the stability of $M$. Therefore, $M_i$ is a stable matching of $I_i(M)$.
\end{proof}

We now consider the special case in which all agents have the same size. By reducing such an \HRS\ instance to an equivalent Hospitals/Residents instance, we obtain the Rural Hospitals Theorem for this setting.

\begin{lemma}
\label{lem:rural_hrs}
For an \HRS\ instance in which all agents have the same size, the following holds.
\begin{itemize}
    \item Every hospital has the same occupancy in every stable matching.
    \item The set of agents matched is invariant across all stable matchings.
    \item Every under-subscribed hospital is matched to the same set of agents in every stable matching.
\end{itemize}
\end{lemma}

\begin{proof}
The high level idea is to construct an equivalent HR instance $I'$ given an $I$, apply the Rural Hospitals Theorem to $I'$, and transfer the properties back to $I$. By applying Rural Hospitals Theorem to $I'$, we know every hospital has the same occupancy, the same set of matched agents, and every under-subscribed hospital is matched to the same set of agents in every stable matching of $I'$. Since we  show that stability and under-subscription coincide between $I$ and $I'$, these three properties transfer directly to $I$.

Let $s$ denote the common size of all agents in $I$. We construct an instance $I' = (A' \cup H', E)$ where for every $a \in A$ we have a corresponding agent $a' \in A'$ with unit size and for every $h \in H$, we have a hospital $h' \in H'$ where $q'(h') =  \lfloor {q(h)/s} \rfloor $. Since every agent has size $s$ in $I$, a hospital $h$ can accommodate a set of agents $S$ in $I$ (i.e., $|S|\cdot s \le q(h)$) iff $h'$ can accommodate the corresponding set $S'$ in $I'$ (i.e., $|S'|\le \lfloor q(h)/s\rfloor$). Therefore, a matching $M$ is feasible with respect to quotas of hospitals and sizes of agents in $I$ if and only if it's corresponding matching $M'$ is feasible in $I'$ (note that $M = M'$ as they are just the same set of edges).

Now we fix a feasible matching $M$ in $I$ and make the following two observations. Firstly, a hospital $h$ has remaining capacity at least $s$ in $I$ iff its corresponding $h'$ has a vacant position in $I'$, so $h$ is under-subscribed w.r.t.\ $M$ in $I$ iff $h'$ is under-subscribed w.r.t.\ $M$ in $I'$. Secondly, by the above observation about accommodating agents, given a pair $(a,h)$, $h$ can displace some set of agents to accommodate $a$ in $I$ iff $h'$ can do so for $a'$ in $I'$. Since preferences are the same, we have that $(a,h)$ blocks $M$ in $I$ iff $(a', h')$ blocks $M$ in $I'$, so $M$ is stable in $I$ iff $M$ is stable in $I'$. This completes the proof.
\end{proof}

\begin{lemma}
\label{lem:induced_instances_equal}
Let $M$ be the matching returned by Algorithm~\ref{algo:gener_ML}, and let $M'$ be any stable matching of the \HRS\ instance with a generalised master list ordering $\langle A_1,A_2,\ldots,A_k\rangle$. Then, for every partition $A_i$, $I_i(M)=I_i(M').$
\end{lemma}

\begin{proof}
We prove the statement by strong induction on partition index $i$.

For $i=1$, there are no higher-preferred partitions. Hence, $I_1(M)=I_1(M')$.

Now assume that, for some $i<k$, $I_j(M)=I_j(M')$ holds for all $j \le i$. By Lemma~\ref{lem:restriction_stable}, the restricted matchings $M_j$ and $M_j'$ are stable matchings of $I_j=I_j(M)=I_j(M')$ for any $j \le i$. Since the agents of $I_j(M)$ are precisely those of $A_j$, they all have the same size. Therefore, by Lemma~\ref{lem:rural_hrs}, every hospital has the same occupancy in $M_j$ and $M_j'$ for all $j\le i$.

Consequently, the same amount of capacity is deducted from every hospital while constructing $I_{i+1}(M)$ and $I_{i+1}(M')$. Hence, $I_{i+1}(M)=I_{i+1}(M')$. This completes the proof.
\end{proof}

\begin{lemma}
\label{lem:rht_for_genML}
Consider an \HRS\ instance with a generalised master list, then 
\begin{itemize}
    \item Every hospital has the same occupancy in every stable matching.
    \item The set of agents matched is invariant across all stable matchings.
\end{itemize} 
\end{lemma}

\begin{proof}
Let $M$ be the matching returned by Algorithm~\ref{algo:gener_ML}, and let $M'$ be any stable matching of the \HRS\ instance $I$. Applying Lemma~\ref{lem:induced_instances_equal} specifically for $i=k$ gives $I_k(M)=I_k(M')=I_k$. Note that $I_k$ has same size agents, and Lemma~\ref{lem:rural_hrs} implies that the occupancy of hospitals is same in both $M_k$ and $M_k'$. Since $A_k$ is the last partition, its immediate that the hospitals have same occupancy in $M$ and $M'$.

By Lemma~\ref{lem:induced_instances_equal}, $I_i(M)=I_i(M')$ for every partition $A_i$. Hence, by Lemma~\ref{lem:restriction_stable}, the restricted matchings $M_i$ and $M_i'$ are stable matchings of the same induced instance $I_i$. Since all agents in $A_i$ have the same size, Lemma~\ref{lem:rural_hrs} implies that the same agents of $A_i$ are matched in $M$ and $M'$. As this holds for every partition, the same agents are matched in $M$ and $M'$. 
\end{proof}

\begin{corollary}
\label{cor:same_size}
Let $M$ and $M'$ be any two stable matchings of an \HRS\ instance with a generalised master list. Then, size of $M$ is equal to size of $M'$.
\end{corollary}
This corollary establishes Theorem~\ref{thm:max_size_gen_master_list}.

\section{\NP-hardness of max-size occupancy-stable matching}
\label{sec:hardness_occ_stable_matching}

In this section, we present a polynomial-time reduction to establish \NP-hardness of the problem of deciding whether an $\mathcal{A}$-perfect occupancy-stable matching exists in an \HRS\ instance, thereby proving Theorem~\ref{thm:occ_stable_hardness}. An $\mathcal{A}$-perfect occupancy-stable matching is an occupancy-stable matching that matches all agents.

In the conference version~\cite{balasundaram_et_al_fsttcs_2025} of our work, we gave a reduction from a variant of the stable marriage problem. In this work, we give a simpler reduction from the Vertex Cover problem which is a well-known \NP-hard problem.
Our new reduced instance also has a master list ordering. As noted earlier, the hardness of max-size occupancy stable matching is implied by
Theorem~5.11 in~\cite{manlove_journal}. The hardness holds because 
\HRS\ is a special case of Course Allocation setting considered in~\cite{manlove_journal}.

Let $G=(V,E)$ be a $3$-regular graph. Let $n = |V|, m = |E|$. Given an integer $k$, the Vertex Cover
problem asks if $G$ admits a vertex cover of size at most $k$.
Using $\langle G,k\rangle$ we construct an \HRS\ instance as follows.

\noindent{\bf Agents. } For every vertex $v_i$, we have an agent $a_i$ with size $3$.
For every edge $e_j$, we have an agent $b_j$ with size $1$.

\noindent{\bf Hospitals. } For every vertex $v_i$, we have a hospital $p_i$ with capacity $3$.
Additionally, we have a hospital $h$ with capacity $3k$. 

Thus, we have $m+n$ many agents and
$n+1$ many hospitals in the reduced instance.
Let the vertices in $V$ be arranged in a fixed but arbitrary order as $v_1, \ldots, v_n$.
Let the edges in $E$ be arranged in a fixed but arbitrary order as $e_1, \ldots, e_m$.
This gives a master list ordering on agents as well as hospitals as follows:

\hspace{2pt}\noindent {\bf Agents: } $b_1 \succ \ldots \succ b_m \succ a_1 \succ \ldots \succ a_n$ \hspace{1.5cm}
\noindent {\bf Hospitals: } $p_1 \succ \ldots \succ p_n \succ h$

The preferences of agents and hospitals are shown below.
For vertex $v_i$, let $e_{i_1}, e_{i_2}, e_{i_3}$ be the three edges incident on $v_i$ in $G$.
For an edge $e_j$ let the two end-points are denoted as $v_{j_1}$ and $v_{j_2}$.
Every agent $a_i$ prefers hospital $p_i$ over $h$.
Every agent $b_j$ prefers the hospitals $p_{j_1}, p_{j_2}$ 
corresponding to vertices $v_{j_1}, v_{j_2}$ in the order that is consistent with the master list ordering.
Every hospital $p_i$ prefers the agents $b_{j_1}, b_{j_2}, b_{j_3}$ corresponding to edges
$e_{j_1}, e_{j_2}, e_{j_3}$ in the order that is consistent with the master list ordering, followed by the agent $a_i$.

\begin{minipage}{0.4\textwidth}
\begin{align*}
    1\leq i \leq n, \ \ \ [3]\ a_i&: p_i \succ h \\
    1\leq j \leq m, \ \ \ [1]\ b_j&: p_{j_1} \succ p_{j_2}
\end{align*}
\end{minipage}
\begin{minipage}{0.4\textwidth}
\begin{align*}
    1\leq i \leq n, \ \ \ [3]\ p_i&: b_{i_1} \succ b_{i_2} \succ b_{i_3} \succ a_i \\
    [3k]\ h&: a_1 \succ a_n
\end{align*}
\end{minipage}

\smallskip
\noindent Let the reduced \HRS\ instance be denoted as $H$.
Next we prove that the reduction is correct.

\begin{lemma}
    If $G$ admits a vertex cover of size at most $k$ then $H$ admits an $\mathcal{A}$-perfect occupancy-stable matching.
\end{lemma}
\begin{proof}
    Let $C$ be a vertex cover in $G$ such that $|C| \leq k$.
    We construct assignment $M$ in $H$ using $C$.
    Consider an edge $e_j$. If both $v_{j_1}$ and $v_{j_2}$ are in $C$ then assign $b_j$ to $p_{j_1}$,
    otherwise there exists a unique $v_x \in \{v_{j_1}, v_{j_2}\} \cap C$, and we assign $b_j$ to the corresponding hospital $p_{x}$.
    For every vertex $v_i$, if $p_i$ has not been assigned with any edge-agent $b_j$ then assign $a_i$
    with $p_i$. Otherwise assign $a_i$ with $h$.

    We show that $M$ does not violate capacities, that is, it's a matching in $H$.
    Since, $C$ is a vertex cover, for each edge $e_j$, $b_j$ is assigned in $M$ to either $p_{j_1}$
    or $p_{j_2}$. Any $p_i$ is either assigned to at least one and at most three $b_j$-agent, each of which
    is size $1$ or to the unique $a_i$ agent which is size $3$.
    Thus $M$ does not violate capacity for any $p_i$. Agent $a_i$ is either assigned to $p_i$ or $h$.
    When agent $a_i$ is assigned to $h$, it implies that $p_i$
    is assigned with at least one $b_j$-agent, implying that $v_i \in C$. Since, $|C| \leq k$ and the size of
    $a_i$ is $3$, $M$ does not violate capacity of $h$.
    By this argument it is also clear that every $a_i$ and every $b_j$ is matched in $M$, thus $M$ is $\mathcal{A}$-perfect.

    Next we show that $M$ is occupancy-stable. By construction of $M$, if $a_i$ is not matched to its rank 1 choice $p_i$ then at least one of $\{b_{i_1}, b_{i_2}, b_{i_3}\}$ is matched to $p_i$. Thus, $p_i$ cannot accommodate $a_i$
    which is its last choice. Thus $(a_i, p_i)$ do not form an occupancy-based blocking pair.
    If $b_j$ is not matched to its rank 1 choice $p_{j_1}$ then it implies that $v_{j_1} \notin C$.
    Therefore, none of the $b$-agents corresponding to edges incident on $v_{j_1}$ is assigned to $p_i$.
    Also, $s(b_j) = 1 < s(a_i) = 3$, therefore none of these $b$-agents form an occupancy-based blocking pair
    with $p_{j_1}$. Thus, $M$ is an occupancy-stable matching.
\end{proof}

\begin{lemma}
    If $H$ admits an $\mathcal{A}$-perfect occupancy-stable matching $M$ then $G$ admits a vertex cover of size at most $k$.
\end{lemma}
\begin{proof}
    Let $C = \{v_i \mid M(a_i) = h\}$. Since the size of $a_i$ is $3$ and capacity of $h$ is $3k$, clearly $|C| \leq k$.
    We show that $C$ is a vertex cover in $G$.
    Suppose not, then there exists an edge $e_j$ such that neither $v_{j_1}$ nor $v_{j_2}$ is in $C$.
    Due to $\mathcal{A}$-perfectness of $M$, we have that $M(a_{j_1}) = p_{j_1}$ and $M(a_{j_2}) = p_{j_2}$.
    This implies that $b_j$ agent is neither matched to $p_{j_1}$ nor matched to $p_{j_2}$ in $M$.
    Note that although $p_{j_1}$ (resp. $p_{j_2}$) prefers $b_j$ over $a_{j_1}$ (resp. $a_{j_2}$), they do not form an occupancy-blocking pair w.r.t. $M$. However, this contradicts that $M$
    is $\mathcal{A}$-perfect. Therefore, $C$ is a vertex cover.
\end{proof}

\smallskip
An $\mathcal{A}$-perfect matching in $H$ has size $m+3n$. Thus we proved that $G$ admits a vertex cover of size at most $k$ if and only if $H$ admits an occupancy-stable matching of size $m+3n$. This establishes Theorem~\ref{thm:occ_stable_hardness}.
\section{NP-hardness for First-Coalition stable matchings in \HRS\ }

\label{sec:hardness_fc}

{In this section, we establish the \NP-hardness of the decision version of the first-coalition stable matching problem. Our reduction is from \CSMTI{} which is defined below. Our reduction shows that the problem remains hard even when all non-unit sized agents have degree $1$.}

The Stable Marriage problem with Incomplete lists (SMI) is the restriction of the \HR{} problem where capacities of all \hospital s are one. Conventionally, the two sets in the Stable Marriage problem are denoted as men and women instead of agents and hospitals. The Stable Marriage problem with Ties and Incomplete lists (\SMTI) is a generalisation of SMI in which preference lists can have ties. The definition of a blocking pair is suitably adapted for tied instances. A pair $(m, w)$ that is not assigned to each other in a matching $M$ blocks $M$ if both $m$ and $w$ strictly prefer each other to their current partners in $M$. A matching $M$ in an \SMTI{} instance is stable if there is no blocking pair with respect to $M$.

Consider a restricted instance of \SMTI{} with an equal number of men and women, in which every woman's preference list is a strict list of length at most three; and each man's preference list is either a strict list of length exactly three or is a single tie of length two. The goal is to decide whether there exists a complete stable matching in such a restricted \SMTI{} instance. This is the  \CSMTI{} problem, which is known to be \NP-hard~{\cite{hrs_hardness}}.

Consider a \CSMTI{} instance $I$, with $n$ men  $\{m_1, m_2, \dots , m_n\}$   and $n$ women $\{w_1, w_2, \dots , w_n\}$. We construct an \HRS{} instance $G = (\A \cup \B, E)$ from $I$ as follows. For each woman $w_i$ we have one \hospital{} $h_i$ with capacity $1$. For each man $m_j$ in $I$ with a preference list which is a tie of length two say $(w_a , w_b)$ where $a < b$,  we create a gadget consisting of twelve \agent s and eight \hospital s as follows. See Fig.~\ref{fig:nonunit_reduction}(i) and (ii).
\begin{itemize}
    \item Agents $\{a_j^1, \ldots, a_j^6\}$ where $\{a_j^1, a_j^2, a_j^3, a_j^4\}$ have unit size  whereas $\{a_j^5, a_j^6\}$ have size $3$.
    \item For each $k = 1, 2$ we have three  agents $\{r_{j,t}^k \ | \ t = 1, 2, 3\}$ where $r_{j,1}^k, r_{j,2}^k$ have unit size  and {$r_{j,3}^k$ has size $2$.}
    \item Two hospitals $h_j^1$ and $h_j^2$ each with capacity $3$.
    \item For each $k = 1, 2$, we have three hospitals $\{p_{j,t}^k \ | \ t = 1, 2, 3\}$ where {$p_{j,3}^k$ has capacity $2$} and the remaining hospitals have capacity $1$.
\end{itemize}
For each man $m_s$ with a strict preference list in $I$, we create a gadget containing $4$ agents and $3$ hospitals. See Fig.~\ref{fig:nonunit_reduction} (iii) and (iv).
\begin{itemize}
    \item A unit sized agent $a_s$. 
    \item Three agents $\{r_{s,t} \ | \ t = 1, 2, 3\}$ where {$r_{s,3}$ has size $2$} and remaining agents are unit sized. 
    \item Three  hospitals $\{p_{s,t} \ | \ t = 1, 2, 3\}$ where {$p_{s,3}$ has capacity $2$} and the remaining hospitals have capacity $1$.
\end{itemize}
This completes the description of the agents and hospitals in $G$. Now we describe the preference lists.
\begin{itemize}
    \item {\bf Preference list for hospital $h_i$ corresponding to woman $w_i$:} Let the preference list of $w_i$ be  $l_{w_i} = m_{i_1}, m_{i_2}, m_{i_3}$. We  start with  preference list of $h_i$ to be set as  $l_{h_i} = l_{w_i}$.  For $t = 1, 2, 3$ the man $m_{i_t}$ in $l_{w_i}$ either has a strict preference list in $I$ or has a single tie of length two. 
\begin{itemize}
    \item If $m_{i_t}$ has a strict preference list, then we denote the man as $m_s$ and replace  $m_s$ in $l_{h_i}$ by the corresponding agent $a_s$.
    \item If $m_{i_t}$ has a tie in its preference list, then we denote the man as $m_j$ and denote the preference list of $m_j$ as $(w_i, w_b)$. If $i < b$, replace $m_j$ with $a_j^1$ in $l_{h_i}$, else with $a_j^2$.
\end{itemize}
This gives us $l_{h_i}$, which is the preference list of hospital $h_i$. 
\item {\bf Preference lists of the vertices in the gadget corresponding to man $m_j$ with a tie:} Recall that the preference list of $m_j$ is a tie of length two, namely $(w_a, w_b)$ where $a< b$. The preferences of the agents and hospitals in the gadget corresponding to $m_j$ are shown in Fig.~\ref{fig:nonunit_reduction}(i) and (ii). Note that the hospitals $h_a$ and $h_b$ corresponding to $w_a$ and $w_b$ appear only in the preferences of $a_j^1$ and $a_j^2$ respectively.
\item {\bf Preference lists of the vertices in the gadget corresponding to man $m_s$ with strict preference list:} Let the preference list of  $m_s$ be  $w_e, w_f, w_g$. Then the preference list of the agent $a_s$ has the corresponding hospitals followed by the gadget hospital $p_{s,1}$. See Fig.~\ref{fig:nonunit_reduction}(iii) and (iv) for the preferences of vertices in the gadget corresponding to $m_s$.
\end{itemize}
\begin{figure}[!ht]
    \centering
    \begin{multicols}{2}

    \begin{minipage}{\linewidth}
    \begin{align*}
        & (1)\ a_j^1: && h_j^1 \succ h_a \succ p_{j,1}^1 \\
        & (1)\ a_j^2: && h_j^2 \succ h_b \succ p_{j,1}^2 \\
        & (1)\ a_j^3: && h_j^2 \succ h_j^1 \\
        & (1)\ a_j^4: && h_j^1 \succ h_j^2 \\
        & (3)\ a_j^5: && h_j^1 \\
        & (3)\ a_j^6: && h_j^2 \\
        & (1)\ r_{j,1}^k: && p_{j,1}^k \succ p_{j,2}^k \succ p_{j,3}^k \\
        & (1)\ r_{j,2}^k: && p_{j,3}^k \succ p_{j,2}^k \\
        & {(2)}\ r_{j,3}^k: && p_{j,3}^k
    \end{align*}
    \vspace{-0.7cm}
    \begin{center}
        \textbf{\RNum{1}}
    \end{center}
    \end{minipage}

    \begin{minipage}{\linewidth}
    \begin{align*}
        & [3]\ h_j^1: && a_j^3 \succ a_j^5 \succ a_j^1 \succ a_j^4 \\
        & [3]\ h_j^2: && a_j^4 \succ a_j^6 \succ a_j^2 \succ a_j^3 \\
        & [1]\ p_{j,1}^k: && a_j^k \succ r_{j,1}^k \\
        & [1]\ p_{j,2}^k: && r_{j,2}^k \succ r_{j,1}^k \\
        & {[2]}\ p_{j,3}^k: && r_{j,1}^k \succ r_{j,3}^k \succ r_{j,2}^k
    \end{align*}
    \vspace{1.5cm}
    \begin{center}
        \textbf{\RNum{2}}
    \end{center}
    \end{minipage}

    \end{multicols}

    \begin{multicols}{2}

    \begin{minipage}{\linewidth}
    \begin{align*}
        & (1)\ a_s: && h_e \succ h_f \succ h_g \succ p_{s,1} \\
        & (1)\ r_{s,1}: && p_{s,1} \succ p_{s,2} \succ p_{s,3} \\
        & (1)\ r_{s,2}: && p_{s,3} \succ p_{s,2} \\
        & {(2)}\ r_{s,3}: && p_{s,3}
    \end{align*}
    \vspace{-0.7cm}
    \begin{center}
        \textbf{\RNum{3}}
    \end{center}
    \end{minipage}

    \begin{minipage}{\linewidth}
    \begin{align*}
        & [1]\ p_{s,1}: && a_s \succ r_{s,1} \\
        & [1]\ p_{s,2}: && r_{s,2} \succ r_{s,1} \\
        & \ p_{s,3}: && r_{s,1} \succ r_{s,3} \succ r_{s,2}
    \end{align*}
    \vspace{-0.25cm}
    \begin{center}
        \textbf{\RNum{4}}
    \end{center}
    \end{minipage}

    \end{multicols}
    \caption{Agents and hospitals corresponding to a man in the SMTI instance}
    \label{fig:nonunit_reduction}
\end{figure}
This completes the description of the reduction. We prove the correctness below.
\begin{lemma}\label{lem:fcpq}
Let $m_j$ be a man with a tie in his preference list in $I$, and let $k \in {1, 2}$. If $G$ admits a first-coalition stable matching $N$, $N$ leaves $(a_{j}^k, p_{j,1}^{k})$ unmatched. Furthermore, $N$ cannot leave $a_j^k$ unmatched.
\end{lemma}

\begin{proof}
Assume there exists $j,k$ as above such that $(p_{j,1}^{k},a_j^k) \in N$. Hence the hospital $p_{j,1}^{k}$ will be removed from $r_{j,1}^k$'s preference list. {Now, the subgraph induced by ${r_{j,1}^k, r_{j,2}^k, r_{j,3}^k, p_{j,2}^k, p_{j,3}^k}$ is exactly the instance in Fig.~\ref{fig:ex1}} and thus does not admit a first-coalition stable matching. Therefore, $N$ cannot be first-coalition stable, a contradiction. The second statement of the lemma follows from the fact that if $a_j^k$ is unmatched in $N$ then $\{a_j^k\}$ with hospital $p_{j, 1}^k$ form a first-blocking coalition with respect to $N$.
\end{proof}

Using similar arguments as above, we get the following lemma.

\begin{lemma}\label{lem:fcs}
Let $m_s$ be a man with a strict preference list in $I$. Then, any matching $N$ in $G$ containing $(a_s, p_{s,1})$ is not first-coalition stable. Furthermore, $N$ cannot leave $a_s$ unmatched.
\end{lemma}

\begin{lemma}\label{lem:fctj}
Let $j$ be such that $m_j$ is a man with a tie in his preference list in $I$. If $G$ admits a first-coalition stable matching $N$ then one of the following holds. Either $T_j^a \subseteq N$ or $T_j^b \subseteq N$, where
\[
T_j^a =
{
(a_j^1,h_a),
(a_j^2,h_j^2),
(a_j^3,h_j^2),
(a_j^4,h_j^2),
(a_j^5,h_j^1)
}
\]
and
\[
T_j^b =
{
(a_j^1,h_j^1),
(a_j^2,h_b),
(a_j^3,h_j^1),
(a_j^4,h_j^1),
(a_j^6,h_j^2)
}.
\]
\end{lemma}
\begin{proof}
Since $N$ is first-coalition stable, from Lemma~\ref{lem:fcpq} the agent $a_j^1$ should be matched to one of ${h_j^1, h_a}$. We divide the proof based on the matched partner of $a_j^1$ in $N$.

\begin{itemize}
\item Assume $(a_j^1, h_a) \in N$. Here we establish that $T_j^a \subseteq N$. In this case, for $N$ to be first-coalition stable, the hospital $h_j^1$ has to be fully subscribed in $N$ with agents higher preferred than $a_j^1$, otherwise {$\{a_j^1\}$ forms a first-blocking coalition with $h_j^1$ w.r.t $N$}. This implies that $(a_j^5, h_j^1) \in N$. In this case the agent $a_j^3$ cannot be left unmatched, {otherwise $\{a_j^1,a_j^3,a_j^4\}$ forms a first-blocking coalition with $h_j^1$}. Thus, $(a_j^3, h_j^2) \in N$. Since $(a_j^3, h_j^2) \in N$, $(a_j^6, h_j^2) \notin N$. The hospital $h_j^1$ is matched to $a_j^5$ and therefore cannot accommodate $a_j^4$. The matching $N$ cannot leave $a_j^4$ unmatched, else $\{a_j^4\}$ with $h_j^2$ forms a first-blocking coalition w.r.t $N$. This implies that $(a_j^4, h_j^2) \in N$. Given the current set of edges in $N$, if $N$ does not match $a_j^2$ to $h_j^2$, then $\{a_j^2\}$ with $h_j^2$ forms a first-blocking coalition w.r.t $N$. Thus $(a_j^2, h_j^2) \in N$. Thus we have proved that if $(a_j^1, h_a) \in N$ then $T_j^a \subseteq N$.

\item Assume $(a_j^1, h_j^1) \in N$. Here, we establish that $T_j^b \subseteq N$. Since $a_j^1$ occupies unit capacity in $h_j^1$, $a_j^5$ cannot occupy $h_j^1$. Hence, $(a_j^5, h_j^1) \notin N$. Furthermore, $(a_j^3, h_j^1) \in N$ else $\{a_j^5\}$ with $h_j^1$ forms a first-blocking coalition w.r.t $N$. If $a_j^4$ is not matched to $h_j^1$, $\{a_j^4\}$ with $h_j^1$ forms a first-blocking coalition w.r.t $N$. Therefore, $(a_j^4, h_j^1) \in N$. Given the current set of edges in $N$, if $a_j^6$ is not matched to $h_j^2$ then $\{a_j^6\}$ with $h_j^2$ forms a first-blocking coalition w.r.t $N$. So, $(a_j^6, h_j^2) \in N$. From Lemma~\ref{lem:fcpq} $a_j^2$ cannot be unmatched and $h_j^2$ cannot accommodate $a_j^2$. Therefore, $(a_j^2, h_b) \in N$. This concludes that $T_j^b \subseteq N$.
\end{itemize}
This completes the proof.

\end{proof}

For each $k\in\{1,2\}$, let
\[
S_j^k=
\{
(r_{j,1}^k,p_{j,1}^k),
(r_{j,2}^k,p_{j,2}^k),
(r_{j,3}^k,p_{j,3}^k)
\}.
\]
Similarly, let
\[
S_s=
\{
(r_{s,1},p_{s,1}),
(r_{s,2},p_{s,2}),
(r_{s,3},p_{s,3})
\}.
\]

\begin{lemma}\label{lem:reducfc}
The instance $I$ admits a complete stable matching if and only if the reduced \HRS{} instance $G$ admits a first-coalition stable matching.
\end{lemma}

\begin{proof}
Consider a complete stable matching $M$ of $I$. Let $G$ be the reduced \HRS{} instance of $I$. Construct a matching $M'$ as follows, starting with $M' = \phi$.
\begin{itemize}
\item For a man $m_j$ with ties in his preference list $(w_a, w_b)$ where $a < b$.
\begin{eqnarray*}
M' = M' \cup \begin{cases}
T_j^a & \text{if } (m_j, w_a) \in M \\
T_j^b & \text{if } (m_j, w_b) \in M
\end{cases}
\end{eqnarray*}
Furthermore for the man $m_j$, we add $M' = M' \cup S_j^1 \cup S_j^2$.
\item For a man $m_s$ with strict preference list in $I$
\begin{eqnarray*}
M' = M' \cup {(a_s, h_i)} \cup S_s.
\end{eqnarray*}
\end{itemize}
\smallskip

It can be verified that $M'$ is feasible in $G$. From Lemmas~\ref{lem:fcpq}, \ref{lem:fcs}, and \ref{lem:fctj}, $M'$ is first-coalition stable.

To prove the other direction, consider a first-coalition stable matching $M'$ of $G$. Construct a matching $M$ in $I$ as follows.
\begin{itemize}
\item For any $j$ such that $m_j$ is a man with ties in his preference list $(w_a, w_b)$
\begin{eqnarray*}
M = M \cup \begin{cases}
{(m_j, w_a)} & \text{if } T_j^a \subseteq M' \\
{(m_j, w_b)} & \text{if } T_j^b \subseteq M'
\end{cases}
\end{eqnarray*}

\item For any $s$ such that $m_s$ is a man with a strict preference list,
\begin{eqnarray*}
    M = M \cup \{(m_s, w_i)\} \ \ \ \mbox{ if $(a_s, h_i) \in M'$}.
\end{eqnarray*}

\end{itemize}
\smallskip

Since the capacity of $h_i$ is $1$, the corresponding woman $w_i$ in $I$ is also matched to at most one man. Every $a_j$ corresponding to a man $m_j$ with ties in his preference list is matched either to $h_a$ or $h_b$, and hence $m_j$ is matched either to $w_a$ or $w_b$ in $I$. By Lemma~\ref{lem:fcs}, $a_s$ has to be matched to one of the hospitals in ${h_e,h_f,h_g}$ in $M'$. Thus, every man with a strict preference list in $I$ is matched to some woman in $I$. Therefore, $M$ is feasible and complete in $I$.

{Since all men with ties in their preference list are matched, they do not block $M$. Suppose there exists a blocking pair $(m_s,w_i)$ for $M$, where $m_s$ is a man with a strict preference list. Then the corresponding agent $\{a_s\}$ forms a first-blocking coalition with $h_i$ w.r.t $M'$ in $G$, leading to a contradiction. Therefore, $M$ is first coalition stable in $I$.}
\end{proof}

Lemma~\ref{lem:reducfc} establishes Theorem~\ref{thm:nonunit_fc_Reduction}.

\vspace{0.1in}
\noindent {\bf Discussion:} In this paper we consider the \HRS\ setting with three stability notions: Stability, first-coalition stability and occupancy stability.
Since classical stability is not guaranteed in general, we investigate it under the generalised master lists on agents and show that under this restriction a stable matching always exists and can be efficiently computed. Moreover, we show  an analogue of the Rural Hospitals Theorem under generalized master lists.

We show that occupancy-stable matchings always exist and can be computed efficiently, but computing a maximum-size one is \NP-hard even under master list orderings. We give an approximation algorithm with guarantee $2+r_{\max}$, strictly better than $3$. Apart from the analysis being a non-trivial improvement over the $3-\frac{1}{s_{\max}}$ guarantee from the conference version, we show that the analysis is tight. A natural direction for future work is to establish approximation hardness for max-size occupancy-stable matchings.

\bibliography{library}

\end{document}